\documentclass[aps,cjp,showpacs,floatfix,twocolumn,superscriptaddress]{revtex4-2}
\usepackage[english]{babel}  
\usepackage{graphicx}
\usepackage{dcolumn}
\usepackage{graphicx}
\usepackage{relsize}
\usepackage{tikz}
\usetikzlibrary{arrows.meta, positioning, shapes.geometric}

\usepackage{xcolor}
\usepackage{float}
\usepackage{cancel}
\usepackage{bm}

\usepackage{tikz}
\usepackage{siunitx}
\usepackage{tablefootnote}
\usepackage{physics}

\usepackage[section]{placeins}
\definecolor{darkblue}{rgb}{0,0,.65}
\usepackage{tikz}
\usepackage[%
 pdfauthor={Benedikt Placke},%
 pdfstartview=FitH,%
 breaklinks=true,%
 bookmarks=true,%
 colorlinks=true,%
 anchorcolor=black,%
 citecolor=red,
 filecolor=black,%
 menucolor=black,%
 urlcolor=blue,%
 linkcolor=red,%
]{hyperref}

\newcommand{\appref}[1]{\hyperref[#1]{App.~\ref*{#1}}}
\usepackage{physics}

\usepackage{bm}
\usepackage{amssymb}
\usepackage{color, soul}
\usepackage{amsmath}
\usepackage{amstext}
\usepackage{latexsym}
\usepackage{graphicx}
\usepackage{mathtools}
\usepackage[ruled]{algorithm2e}
\begin{document}

\title{Local supersolid in moiré modulated Bose-Hubbard model using  density-matrix renormalization group method }

\author{Siyu Xie}

\affiliation{College of Physics, Taiyuan University of Technology, Shanxi 030024, China}
 \author{Qiang Xu}

\affiliation{College of Physics, Taiyuan University of Technology, Shanxi 030024, China}

\author{Qianqian Shi}
\email{shiqianqian@cqu.edu.cn}
\affiliation{Centre for Modern Physics, Chongqing University, Chongqing 400044, The People’s Republic of China}

\author{Wanzhou Zhang}
\email{zhangwanzhou@tyut.edu.cn}
\affiliation{College of Physics, Taiyuan University of Technology, Shanxi 030024, China}

\date{\today}

\begin{abstract}
The search and characterization of supersolid phases remain a central topic in condensed matter physics. Inspired by the experimental discovery of local superfluid and insulating phases in two-dimensional moiré optical lattices [Meng et al., Nature 615, 231 (2023)], we systematically explore the emergence of a \textit{local supersolid} ($l$SS) phase in a one-dimensional Bose-Hubbard model subjected to a moiré potential, using the density-matrix renormalization group method. We impose a maximum site occupation $n_{\rm max}=2$ to realize the soft-core boson constraint. In the absence of nearest-neighbor repulsion, we identify the conventional superfluid,  local superfluid, Mott insulator, and moiré-induced insulator phases. When the nearest-neighbor repulsion  is turned on, the $l$SS phase emerges in the strong-moiré regime. This phase is uniquely characterized by three key signatures: (i) coexisting local staggered density order and local off-diagonal coherence within isolated moiré supercells; (ii) exponentially decaying global off-diagonal correlations; and (iii) a vanishing global structure factor in the thermodynamic limit, while the local structure factor remains finite. These features clearly distinguish the $l$SS from the conventional global supersolid (SS) phase, which exhibits algebraic correlations and a finite global structure factor.  Our results provide a complete microscopic picture of local quantum phases in moiré lattices and offer clear experimental observables for detecting $l$SS states with ultracold atoms.

\end{abstract}

\maketitle

\section{Introduction}
\label{sec:model}

The supersolid (SS) phase
is a novel quantum phase where diagonal and off-diagonal long-range orders coexist, and relevant research has been extensively conducted from both theoretical and experimental perspectives~\cite{solid_sf,3} .
The theoretical exploration of the SS phase has been carried out in multiple physical systems and lattice models, with the formation of the phase found to be highly dependent on lattice geometry, boson interaction characteristics and interaction range~\cite{ss1,ss2,ss3,Triangle2}.
Theoretically, one of the most common models for describing supersolids is the extended Bose-Hubbard (BH) model \cite{Jaksch1998}.
Specifically, the extended BH model is an extension of the standard BH model, incorporating nearest-neighbor interactions. When dipolar excitons are employed as the bosonic particles, this model has also been successfully proposed theoretically~\cite{ebhexci}.
For hard-core bosons with nearest-neighbor interactions, early numerical simulations demonstrated that the SS phase cannot be realized in square~\cite{sqhard, Hebert,Sengupta}, honeycomb~\cite{honey}, kagome~\cite{ka}, star~\cite{star}, and Shastry-Sutherland lattices~\cite{ss}, due to the intrinsic instability of the SS phase~\cite{Sengupta,sqhard}, which leads to phase separation into a solid phase and a superfluid phase for all interaction strengths.

Nevertheless, this situation can be overcome by introducing geometric frustration in triangular lattices, which stabilizes the SS phase~\cite{ss2,ss3,Triangle2}.
Specifically, the triangular lattice geometry inherently induces frustration in the nearest-neighbor repulsion—an effect that gives rise to a classical ground state with massive macroscopic degeneracy. When quantum hopping is introduced, quantum fluctuations lift this degeneracy, driving the system to resolve the classical frustration: it selects a state that minimizes kinetic energy via particle delocalization (superfluidity) while simultaneously preserving broken translational symmetry (solid order).
Moreover, in the soft BH model, doping bosons or introducing vacancies in bipartite lattices can also lead to the formation of the SS phase.
Further theoretical studies have demonstrated that dipole-dipole interactions and other long-range interactions can also stabilize the SS phase~\cite{49,51,47,48,55,50,43,45,41,58,46}.

In contrast to the rich theoretical predictions, the experimental realization and verification of the SS phase have experienced great challenges, and the research progress varies significantly across different systems. The realization of the SS phase in ~$^4\text{He}$~\cite{59, 60, 64} systems has proven extremely difficult in experiments, and no stable SS phase has been successfully observed so far. Major experimental breakthroughs have been achieved in realizing the supersolid (SS) phase using ultracold atomic systems in optical lattices~\cite{ss2017-1,ss2017-2,ss2017-3,Chomaz2019,Bottcher2019}.
These include realizations via cavity-mediated interactions, spin-orbit-coupled quantum gases, and long-range dipolar interactions, all providing direct evidence for supersolidity.
In 2022, the extended Bose-Hubbard model was experimentally implemented using dipolar excitons~\cite{eBH_Lagoin2022}.
More recently, the scope of supersolid research has been rapidly expanded to photonic-crystal polariton condensates~\cite{Trypogeorgos2025} and two-dimensional dipolar quantum gases~\cite{he2025observationsupersolidstripestate}, further enriching the landscape of accessible quantum phases.
While these platforms exhibit rich many-body phenomena, they do not directly implement the extended BH model with ultracold atoms, which remains a key motivation for the present theoretical study.
In addition to cold atom and exciton systems, the SS phase has also been identified in realistic spin materials, such as Na$_2$BaCo(PO$_4$)$_2$~\cite{56} and K$_2$Co(SeO$_3$)$_2$~\cite{zhu2025}, expanding the experimental platform for SS phase research. Furthermore, in driven quantum gas systems, sound modes with characteristics analogous to the SS phase have been observed~\cite{Liebster2025,Trypogeorgos2025}, and these unique dynamic modes can be used as an important probe to characterize the dynamic and far-from-equilibrium quantum states related to the SS phase.

With the development of research on the SS phase, a new platform would be of great value for understanding the characteristic features of the SS phase. In recent years, the realization of the moiré potential~\cite{Meng_2023} adds some new degrees of freedom to the cold atom on the optical lattices, which has the potential to lead to new physics related to the SS phase.

\begin{figure}[t]
\centering
\includegraphics[width=1\linewidth]{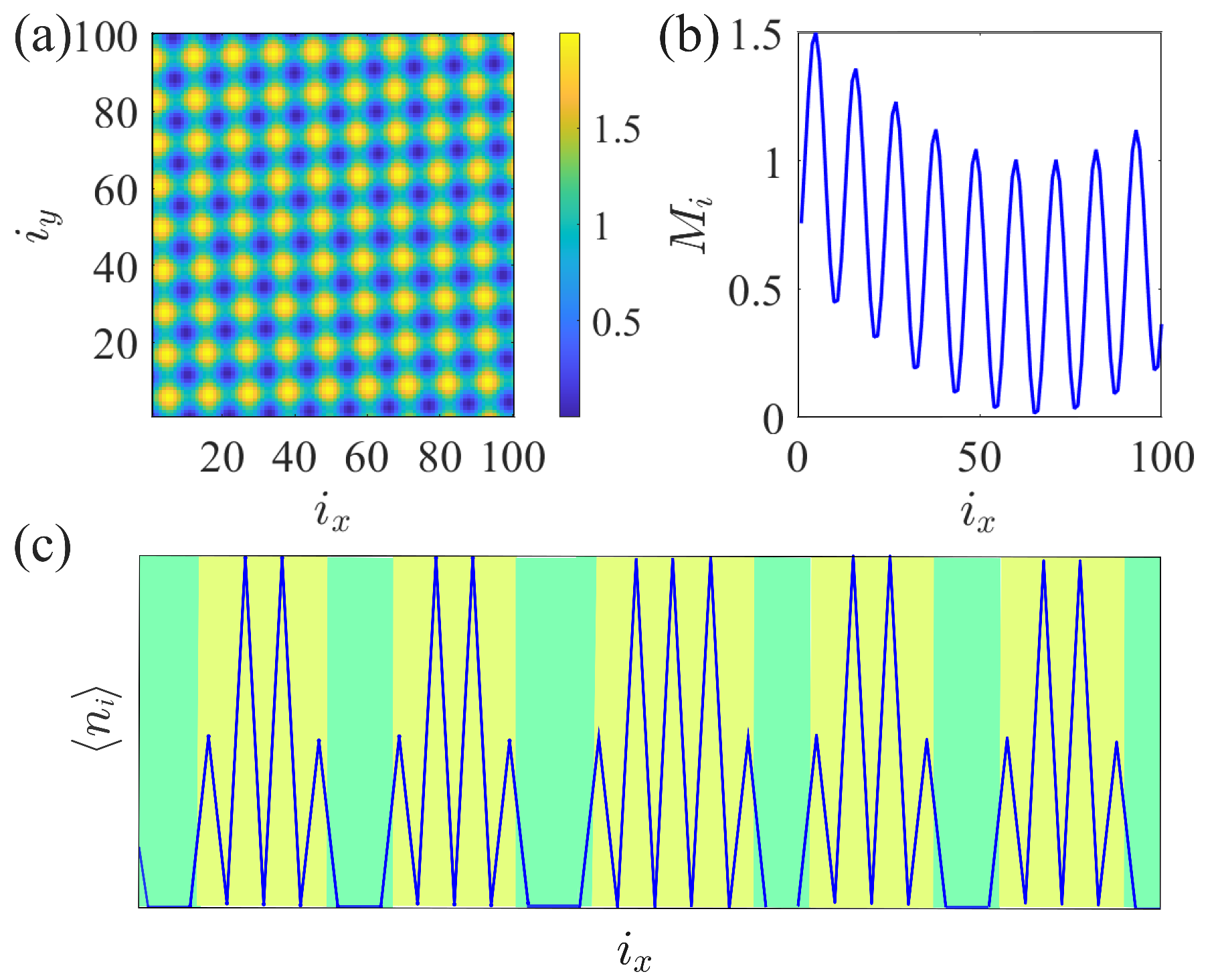}
\caption{
Schematic diagram of the moiré potential in a bosonic lattice system.
(a) Two-dimensional spatial distribution of the moiré potential $M_R$.
(b) One-dimensional cross-section along $i_x$, showing alternating potential wells and barriers.
(c) One-dimensional staggered particle density $\langle n \rangle$.
The local supersolid phase appears in yellow potential wells with non-integer filling,
while green regions act as impenetrable barriers.
It is characterized by coexisting staggered order, finite compressibility, and local off-diagonal correlations.
}

\label{fig:lattice}
\end{figure}

A moiré potential is formed by the superposition of two periodic potentials with different wavelengths. When the wavelength ratio of these two periodic potentials is an irrational number, they are incommensurate; in this case, the resulting moiré potential is a type of incommensurate potential and exhibits quasiperiodicity~\cite{morie_PRL2021}. An example of a moiré potential is shown in Figs.~\ref{fig:lattice} (a) and (b) for both two-dimensional and one-dimensional lattices. The non-uniform potential can result in  insulating (I) and local superfluid (SFII) phase, which has also been observed in experiments~\cite{Meng_2023} and simulations based on the Gross-Pitaevskii equation~\cite{62,63}.
For example, in the non-uniform insulator, the number of particles at the bottom of the local potential wells is greater than that at barriers. In the SFII phase, the potential barrier regions are in the insulating phase (e.g., the empty phase), while the potential well regions are local superfluid regions.
In fact, the spatial extent of local regions is not defined by a fixed number of lattice sites. The localization phenomenon is determined by the confinement of moiré potential wells and potential barriers. The width of these  domains is not fixed, but depends on the chemical potential, the amplitude of the moiré potential, and the local spatial position within the system.

Inspired by experimental observations of SFII and non-uniform insulator I phases under moiré potential,
one may ask {\it whether $l$SS phenomena can emerge when introducing nearest-neighbor repulsive interactions? }
Under moiré potential modulation, the one dimenisional (1D) $l$SS phase is a coexistence of local solid and superfluid order: the latter is characterized by the nearest-neighbor tunneling expectation value \(\langle a_{i}^{\dagger} a_{i+1}\rangle\), while the former appears as a staggered particle number density distribution (Fig.~\ref{fig:lattice} (c)). Unlike the global SS phase, 1D lSS has a vanishing global structure factor (at peaks, thermodynamic limit) and exponentially decaying off-diagonal global correlations; the global SS, by contrast, has a non-zero structure factor and algebraically decaying off-diagonal correlations.

In this manuscript, we study the BH model with moiré potential using the density-matrix renormalization group(DMRG) method~\cite{dmrg}. We mainly focus on the 1D lattice with and without nearest-repulsion $V$, and we  reveal many local phases induced by large  moiré potential.  These interesting properties can be experimentally observed by the quantum gas microscope~\cite{Bakr2009,Gross2021}.

The outline of this paper is as follows. Sec.~\ref{sec:ham_method} presents the model Hamiltonian and numerical details of the DMRG method. Sec.~\ref{sec:review} summarizes and reviews the key local quantum phases. Sec.~\ref{sec:diagram} shows the complete phase diagrams and detailed physical analysis. Finally, Sec.~\ref{sec:con} provides the discussion and conclusions. In the Appendix, we present additional parameter calculations based on Bloch wave function analysis.

\section{Model and Method}
\label{sec:ham_method}
\subsection{The extended BH model with moir\'e potential}
\label{sec:model}
The extended BH model can be written as~\cite{Jaksch1998,Fisher1989Boson,20,Bloch2008Many},

\begin{align}
 H_0 &= -t \sum_{\langle i,j\rangle}(a_{i}^{\dagger} a_{j}+a_{j}^{\dagger} a_{i})  +\frac{U}{2} \sum_{i} n_{i}\left(n_{i}-1\right) \notag \\
 &+ V \sum_{\langle i j\rangle} n_{i} n_{j} -\mu \sum_i n_i,
\end{align}
which contains four parameters: tunneling amplitude between neighboring sites $t$, on-site many-body interaction energy $U$, nearest-neighbor repulsive interaction $V$, and chemical potential $\mu$.  In this model,  $\hat{a}^\dagger$ and $\hat{a}$ are the creation and annihilation operators, respectively. $\hat{n}_i = \hat{a}_i^\dagger \hat{a}_i$ is the number operator at site $i$.
When $V=0$, the equation would be reduced to the standard BH model. When $V\ne 0$,
 the phase diagram of this model  $H_0$ within  the plane has been investigated in plane ($t/V$, $\mu/t$).

Following the experiments by Meng {\it et. al.} in Ref. \cite{Meng_2023},  the hopping amplitude $t$ is controlled by the depth of the optical lattice $V_0$, which can be estimated as
\begin{equation}
t = \frac{4}{\sqrt{\pi}} E_R \left( \frac{V_0}{E_R} \right)^{3/4} e^{-2\sqrt{V_0/E_R}}.
\label{eq:hopping}
\end{equation}
The local repulsion $U$ depends on both the lattice depth and the $s$-wave scattering length $a_s$, and is given by~\cite{zwerger2003}
\begin{equation}
U = g\int |w(x)|^4 dx=\sqrt{\frac{8}{\pi}} k a_s E_R \left( \frac{V_0}{E_R} \right)^{3/4},
\end{equation}
where $w(x)$ is the Wannier function, $E_R = \hbar^2 k^2/(2m)$ is the recoil energy and $k = \pi/d$ is the lattice wave vector.
Although the nearest-neighbor repulsion $V$  is typically weak in conventional cold atomic systems as it originates from the weak overlap of Wannier functions on adjacent sites,
it can be significant and widely tuned in dipolar systems~\cite{ss2017-3,Chomaz2019,Chomaz2019,Bottcher2019}
or excitonic systems~\cite{eBH_Lagoin2022} as realized in recent experiments. Here we focus on the theoretical many-body physics driven by the various parameters.

In Ref.~\cite{Meng_2023}, the extended BH model with moir\'e potential can be written as
\begin{equation}
 H =
H_0  + \sum_{i} M_in_{i},\label{eq:ham1}
\end{equation}
 where the site-dependent incommensurate chemical potential $M_i$ can be written as
\begin{equation}
M_{i}=M_{R}\left(\sin ^{2}(\phi_i) + \sin^2(\phi_i^{\perp}) \right), \label{eq:mi}
\end{equation}
where $\phi_i = i_{x} \pi \cos \theta+i_{y} \pi \sin \theta$, $\phi_i^\perp = i_{y} \pi \cos \theta - i_{x} \pi \sin \theta$, with $i = (i_x, i_y)$. The subindex $i_{x}$ and $i_{y}$ label the position of the $i$-th site in the two-dimensional space.
Without loss of generality, only the typical angle $\theta = 5.21^\circ$ is chosen ~\cite{Meng_2023}, the theoretical derivations and implications from other values of the angle in later sections are discussed in later section.
The parameter $M_R$ represents the  moir\'e potential  strength,  which is related  to the microwave Rabi frequency
between the two spin states atoms, which act as the two synthetic layers.
It can be independently and continuously tuned over a wide range
(from 0 up to $1\,E_R$) by adjusting the power of the microwave field,
without affecting the tunneling $t$ or on-site interaction $U$.

In this paper, we focus on the soft-core BH model, where the maximum occupation for each site is fixed to $n_{max}=2$. Such a constraint can be realized via a three-body restriction mechanism~\cite{zhang2013_pair,Zhang2017,Keilmann2011,greschner2015anyon}.

\subsection{The numerical method (DMRG)}

We solve Eq.~\ref{eq:ham1} using the DMRG method ~\cite{dmrg, 6,36}, which can be easily implemented in the ITensor library
~\cite{itensor}. This method is efficient in dealing with ground state and the phase diagram of the strongly correlated 1d quantum
systems with very high precision~\cite{dmrg_review}. The periodic boundary condition is used to eliminate the finite-size effect.
While the full lattice is treated with periodic boundary conditions, the local region  is characterized on a truncated sub-region of the whole chain. Consequently, this local segment effectively exhibits open boundary behavior.
The details for the DMRG implementations are specified as follows. The number of sweeps ranges from 50 to 180, and a strict truncation error threshold is set to as low as \(1 \times 10^{-20}\).
The bond dimension of matrix product states(MPS) varies from 50 to 500. Moreover, the advanced noise  parameter is also added as \(1 \times 10^{-5}\). The noise term~\cite{noise}  can improve the efficiency of the MPS method in the solid or insulator phases, where quantum fluctuations are absent, leading to reduced DMRG efficiency. By introducing noise to perturb the density matrix, it enables the construction of a new MPS basis that better captures correlations, which are missed in the unperturbed state. In our simulations, we adopt two schemes, with and without noise, to ensure the consistency and reliability of the results.
Our partial data have also been compared with those obtained via the directed-loop stochastic series expansion  quantum Monte Carlo method~\cite{directed_loop} and exact diagonalization method, and good consistency is achieved.

\subsection{Measurable physical quantities}
\label{sec:quantities }

\begin{itemize}
    \item Particle density and local superfluid order parameter : $\langle n_{i}\rangle$ and $\langle a_{i}\rangle$
    \item  Compressibility $\kappa$:
\begin{equation}
  \kappa=L \left(\left\langle n^{2}\right\rangle-\langle n\rangle^{2}\right),
\end{equation}
where $L$ denotes the total number of lattice sites. The compressibility $\kappa$ serves as a critical indicator to distinguish between compressible phases, such as SF, SFII, $l$SS and SS, where $\kappa > 0$ and incompressible phases, such as MI phase and the moir\'e-induced I phase where $\kappa \to 0$.
    \item Off-diagonal correlation function:
    \begin{equation}
        C(r)=\langle a_{i}^{\dagger} a_{i+r}\rangle.
    \end{equation}
In the superfluid (SF) or SS phase, the correlation decays algebraically, which means the system exhibits
an off-diagonal quasi-long-range order. In contrast, in the density wave (DW) phases, the correlaiton tends to decay exponentially.
    \item The rescaled structure factor~\cite{ss1,ss2,ss3,Triangle2,3,49,51,47,48,55,50,43,45,41,58,46}:
    \begin{align}
        S({k})/L=\frac{1}{L^{2}} \sum_{i, j=1}^L  e^{i {k} \cdot({r}_{i}-{r}_{j})}\langle n_{i}n_{j}\rangle,
    \end{align}
    where ${k}$ is the wave vector, and ${r}_{i}$, ${r}_{j}$ are the positions of lattice sites $i$ and $j$, respectively.
     The structure factor $S({k})$  is scaled by $L$ to define an  order parameter that remains finite in the thermodynamic limit for solid phases. For this solid phase with a   lattice configuration such as $1010\cdots$, the structure factor exhibits a non-zero value at $k= 2\pi/2=\pi$. The factor 2 in the denominator represents the length of the translation-invariant vector for particle distribution.
 The structure factor $S(k)$ characterizes the diagonal long-range order of the system. A finite peak in $S(k)/L$ (as $L \to \infty$) indicates the spontaneous breaking of translational symmetry, which is the signal of solid and SS phases.
     \item  Local structure factor:
    \begin{align}
        S_c( {k})/L_c=\frac{1}{L_c^{2}} \sum_{i, j=1}^{L_c}  e^{i {k} \cdot({r}_{i}-{r}_{j})}\langle n_{i}n_{j}\rangle,
    \end{align}
    where ${k}$ is the wave vector, and ${r}_{i}$, ${r}_{j}$ are the positions of lattice sites $i$ and $j$, respectively.
    The density-density correlation in the local structure factor  $S_c({k})/L$ is confined only to the local region we select, such as the particle correlation $\langle n_{i}n_{j}\rangle$ between two dashed vertical lines, and the subscript ``c" is the abbreviation of ``cut".
   It allows us to detect the robust crystalline order that survives within local supercells (as seen in the $l$SS phase), even when the global structure factor vanishes.
    \item Momentum distribution function:
    \begin{equation}
        n({k})=\frac{1}{L} \sum_{i j} e^{i {k} \cdot\left({r}_{i}-{r}_{j}\right)}\left\langle a_{i}^{\dagger} a_{j}\right\rangle.
    \end{equation}
 The momentum distribution function $n(\mathbf{k})$ describes the population of atoms at different momentum states, which directly reflects the coherence and correlation properties of the system.
    Sharp peaks in $n(\mathbf{k})$ indicate long-range phase coherence, while a broad and flat distribution represents short-range or localized states.
    The sharpness of the momentum distribution varies significantly in different phases and can be directly observed in experiments~\cite{Kato2008,Ernst2010,Guo2024}.
    \item Visibility~\cite{nu}:
    \begin{equation}
        v=\frac{\max ({n}({k}))-\min ({n}({k}))}{\max ({n}({k}))+\min ({n}({k}))}.
    \end{equation}
    Visibility measures the interference pattern  contrast in the momentum distribution~\cite{nu}. A high visibility means that there is a clear difference between the maximum and minimum values of the distribution, suggesting coherent behavior, and thus a superfluid order.
\end{itemize}

\begin{figure}[tb]
\includegraphics[width=0.9\linewidth]{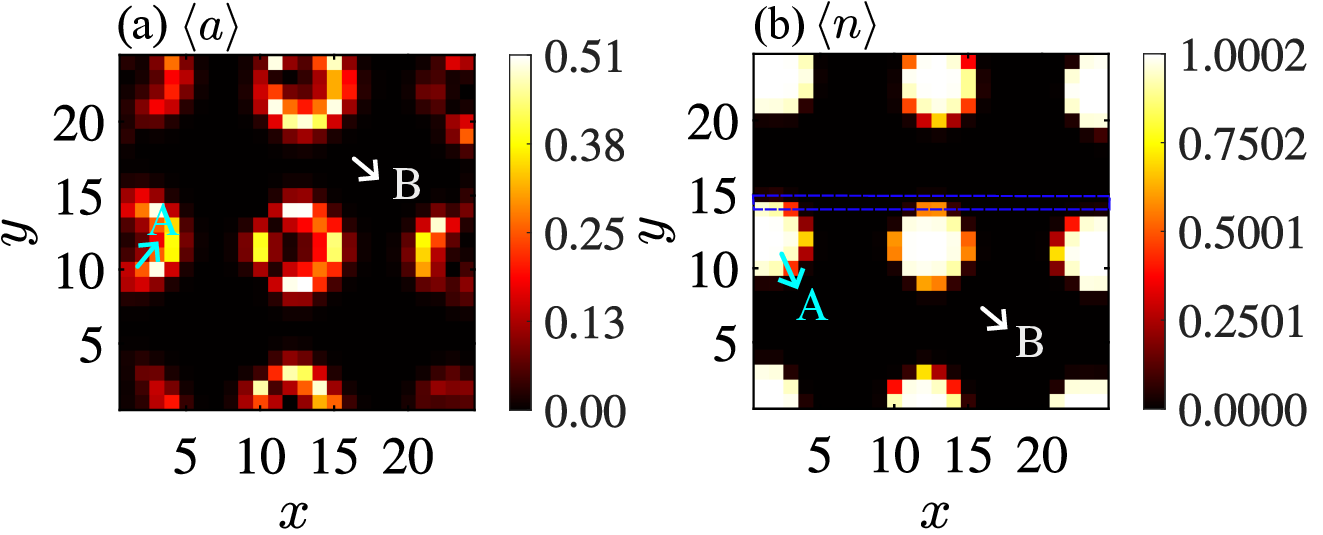}
\caption{
Real-space configurations of (a) superfluid order parameter $\langle a_i \rangle$  and (b) particle density $\langle n_i \rangle$  for the local superfluid (SFII) phase (DMRG calculations).
Color bars quantify $\langle a_i \rangle$ (0–0.51) and $\langle n_i \rangle$ (0–1.0002).
``A" (bright) and ``B" (dark) regions denote moiré potential wells/barriers, respectively.
Island-like $\langle a_i \rangle$ distribution confirms SFII phase: superfluid coherence is confined to wells and does not percolate globally.
}
\label{fig:t=0.5U1mu0.5an}
\end{figure}

\section{Review of the phases}
\label{sec:review}
\subsection{The typical phases}

In Ref.~\cite{Meng_2023}, the phase diagram comprises four distinct phases in moiré lattice systems.
Benefiting from the rapid progress in twisted bilayer optical lattices, moiré quasicrystals and periodically modulated trapping potentials, rich quantum phases of ultracold bosons have been widely explored in both theoretical~\cite{Zhang2025PRB,Ding2025PRA,Zeng2025} and experimental~\cite{science.adh3023} studies in recent years.

Below, we classify the ground-state phases and their characteristic features.

(I) \textit{Uniform MI phase}: The simplest phase, where the particle number per lattice site is an integer with no fluctuations, is indicated  by  $\langle a_i\rangle=0$ and $\langle n_i\rangle=\text{integer}$ ($i=1,\cdots,L$).
This phase already exists in the standard BH model without moiré modulation or under very weak moiré potentials~\cite{Jaksch1998,greiner2002quantum}.

(II) \textit{Non-uniform insulator phase}:
With larger $M_R$, disk-shaped domains of density 2 emerge against a background density of 1~\cite{Meng_2023}. A cross-section along any direction shows $\langle n_i \rangle$ follows the pattern (111 2222 1111 22222…), forming an alternating sequence of 1s and 2s. This phase was also be  named as the I phase~\cite{Meng_2023}.
Similar phase-separated insulating domains induced by moiré potential modulation have also been reported in twisted bilayer bosonic systems, where periodic insulating islands are surrounded by low-density backgrounds~\cite{Zhang2025PRB}.

(III) \textit{Local SF phase} (named SFII phase in Ref.~\cite{Meng_2023}):

In Fig.~\ref{fig:t=0.5U1mu0.5an}, we present typical real-space configurations of the order parameters obtained by DMRG  with numerical values indicated by the corresponding colorbars.
Figure \ref{fig:t=0.5U1mu0.5an}(a) displays the spatial distribution of the local superfluid order parameter $\langle a_i \rangle$. The yellow regions in this panel exhibit high coherence with $\langle a_i \rangle \approx 0.3 \sim 0.5$, corresponding to potential wells. In contrast, the black regions show negligible coherence with $\langle a_i \rangle \approx 0$, which correspond to potential barriers and the intermediate  region between barriers and wells.
Such spatially fragmented superfluidity is a typical property  of moiré and quasicrystal lattices, where repulsive interactions and quasiperiodic modulation jointly confine superfluid coherence into isolated domains~\cite{Ding2025PRA,Fan2026}
As characterized via percolation in Refs.~\cite{Johnstone_2021,Ding2025PRA}, $\langle a_i \rangle$ is mapped to a site percolation configuration.
Figure \ref{fig:t=0.5U1mu0.5an}(b) shows the spatial distribution of particle density $\langle n_i \rangle$.

Both $\langle a_i\rangle$ and $\langle n_i \rangle$ exhibit island-like spatial distributions, with particle coherence confined within the length of moiré superlattice unit cells. As characterized via percolation in Refs.~\cite{Johnstone_2021,Ding2025PRA}, $\langle a_i\rangle$ is mapped to a site percolation configuration using a threshold $\epsilon\approx 10^{-2}$: sites with $\langle a_i\rangle>\epsilon$ (yellow area, labeled as ``A") are occupied, while others (black area, labeled as ``B") are not. Since yellow regions do not span the entire lattice, this phase is termed the local SF phase. The density distribution shown in panel (b) further clarifies the spatial variation of $\langle a_i \rangle$: black regions with $\langle a_i \rangle \approx 0$ in panel (a) correspond to integer-filling domains ($\langle n_i\rangle = 0$ or $1$), which are governed by the non-uniform  term $-(\mu - M_i) n_i$.
Similar isolated superfluid islands embedded in insulating backgrounds have been uncovered in interaction-induced moiré lattices and twisted bilayer optical lattices~\cite{Zhang2025PRB,Zeng2025}.

(IV) \textit{Global SF phase}: With stronger quantum fluctuations, sites with $\langle a_i\rangle\neq0$ connect and span the entire lattice. Despite non-uniform spatial distributions of physical quantities (due to moiré potential modulation), this phase is still referred to as the SF phase for simplicity.

\begin{table}[t]
\centering
\caption{Summary of order parameters and phase properties.
``al.'' and ``ex.'' denote algebraic and exponential decay of correlations $C(r)$, respectively;
while ``int.''=integer, ``uni.''=uniform, ``non-uni.''=non-uniform.
The $l$SS phase is characterized by a \emph{local} nonzero structure factor $S(k)$ within isolated regions, rather than a global one.}
\setlength{\tabcolsep}{3pt}
\resizebox{\linewidth}{!}{%
\begin{tabular}{lccccc}
\hline\hline
Phase & $\langle n_i \rangle$ & $C(r)$ & $S(k_{\text{max}})/L$ & $\kappa$ & Nature \\
\hline
SF    & non-uni.              & al. & $0$ & $\neq 0$ & global  \\
MI    & uni. int.             & ex. & $0$ & $= 0$    & insulator \\
SS    & staggered non-int.    & al. & $\neq 0$ & $\neq 0$ & global \\
SFII  & non-uni.              & ex. & $0$ & $\neq 0$ & local SF \\
I     & non-uni. int.         & ex. & $0$ & $= 0$ & incompressible \\
\hline
$l$SS & staggered & ex. & local: $\neq 0$ & $\neq 0$ & local \\
\hline\hline
\end{tabular}%
}
\label{tab:phases}
\end{table}

\begin{figure}[b]
\includegraphics[width=0.45\textwidth]{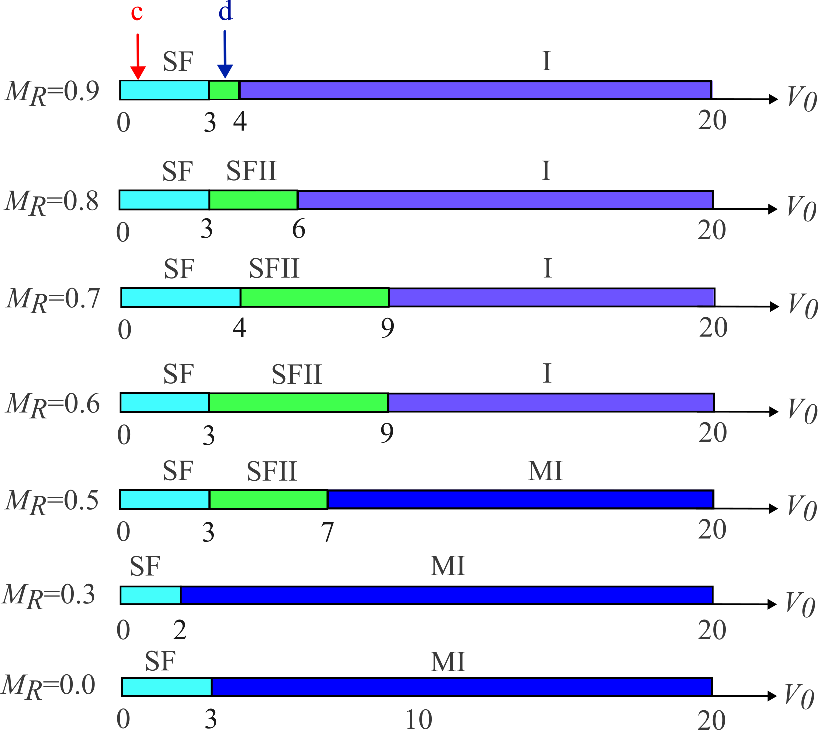}
\caption{A schematic phase diagram of the extended BH model in a one-dimensional moir\'e potential, computed at fixed parameters $U=1$, $\mu=0.6$, and $V=0$. The horizontal axis represents depth of the potential well $V_0$, while the vertical axis corresponds to the moir\'e modulation ratio $M_R$, ranging from $0$ to $0.9$. At small $V_0$, the system is in the SF phase. As $V_0$ increases, the system directly transitions to the MI phase in the low-$M_R$ regime. In the high-$M_R$ regime, by contrast, the system first enters the SFII phase before eventually reaching the insulating I phase at large $V_0$.}
\label{fig:v0_xiangtu}
\end{figure}

\subsection{Identification of the phases}

Table \ref{tab:phases} summarizes the key order parameters and physical properties of various quantum phases, including the local particle density $\langle n_i \rangle$, correlation function $C(r)$, reduced structure factor $S(k)/L$, and the corresponding phase nature. With small morie potential, the global SF phase is characterized by a non-uniform density, algebraic decay of the correlation function, no peaks in the structure factor, and represents a globally coherent superfluid state. In contrast, the local SFII  phase  exhibits a non-uniform density and exponential decay of correlations, with no visible peaks in the structure factor. The MI phase shows an integer filling density, exponential correlation decay, and a zero structure factor, behaving as a typical insulator. The incompressible I phase  is identified by a non-uniform integer density and exponential correlation decay. The SS phase  features a non-uniform density, algebraic correlation decay, and finite structure factor peaks, hosting the coexistence of superfluid and solid orders. Finally, the local $l$SS phase possesses a staggered density distribution, exponential correlation decay, zero global but nonzero local structure factor peaks, and corresponds to the coexistence of local superfluidity and solidity.

The difference between the SFII phase and the SS phase can be directly observed from their configurations.
From the perspective of particle number density, the particle number density of the SFII phase exhibits no staggering, whereas the SS phase shows a staggered particle number density.
Regarding the decay behavior of the off-diagnoal correlation function $C(r)$: the former decays exponentially, while the latter decays algebraically.
In terms of the structure factor, the SFII phase exhibits no peak near $k=\pi$, while the other phase shows a clear peak.

The easiest way to identify the presence of a $l$SS phase is to inspect its particle density profile.
If isolated density islands emerge in the particle number density distribution, and these islands exhibit density staggering with non-integer staggered occupation such  $0~0~0.15~0.85~0.15~0.85\cdots0~0$, the system corresponds to a $l$SS phase. The staggered density inside the islands indicates that the system possesses local solid order.
These two fundamental orders can be separately quantified using dedicated physical probes: local solid order is captured by the local structure factor $S_c(k)/L_c$, while local superfluid order is characterized by the off-diagonal correlation $C(1)$.
In fact, it has been confirmed that the global structure factor $S(k)/L$ vanishes in the thermodynamic limit and global correlations decay exponentially, which rules out the existence of a globally ordered supersolid.

\section{results}
\label{sec:diagram}

The reasons for exploring atoms in 1D lattices with a moiré potential are as follows.  For numerical methods such as the DMRG method, results for 1D systems are more accurate, converge more quickly, and require less computational cost.
Experimentally, the physics of the 1D morie lattice has been realized~\cite{1d,Meng_2023}.
1D moiré potential $V(x)$ is obtained by fixing the 2D potential $V(x,y)$  at $y=y_0$ without loss of generality.
As shown in Fig. \ref{fig:t=0.5U1mu0.5an},
   we select the line $y_0 = 14$.

\subsection{Local superfluid and non-uniform insulator in a non-interacting system ($V=0$)}

Figure \ref{fig:v0_xiangtu}  presents a schematic phase diagram that includes four phases: SF, SFII, MI, and I.
The range of phases is represented by line segments in different colors.
If $M_R=0$, the system undergoes the SF-MI transitions at $V_0=3$.
The density of the system is uniform in the SF phase.
When the moiré potential $M_R$ is small but no-zero, i.e. $M_R=0.2$ the system  still exhibits a SF-MI phase, but at $V_0=2$. As $M_R$ increases, such as $M_R=0.7$, the SF evolves into an SFII phase at deeper potentials ($V_0=4$) and is further transformed into an I phase at deeper potentials ($V_0=9$).

\begin{figure}[b]
\includegraphics[width=0.5\textwidth]{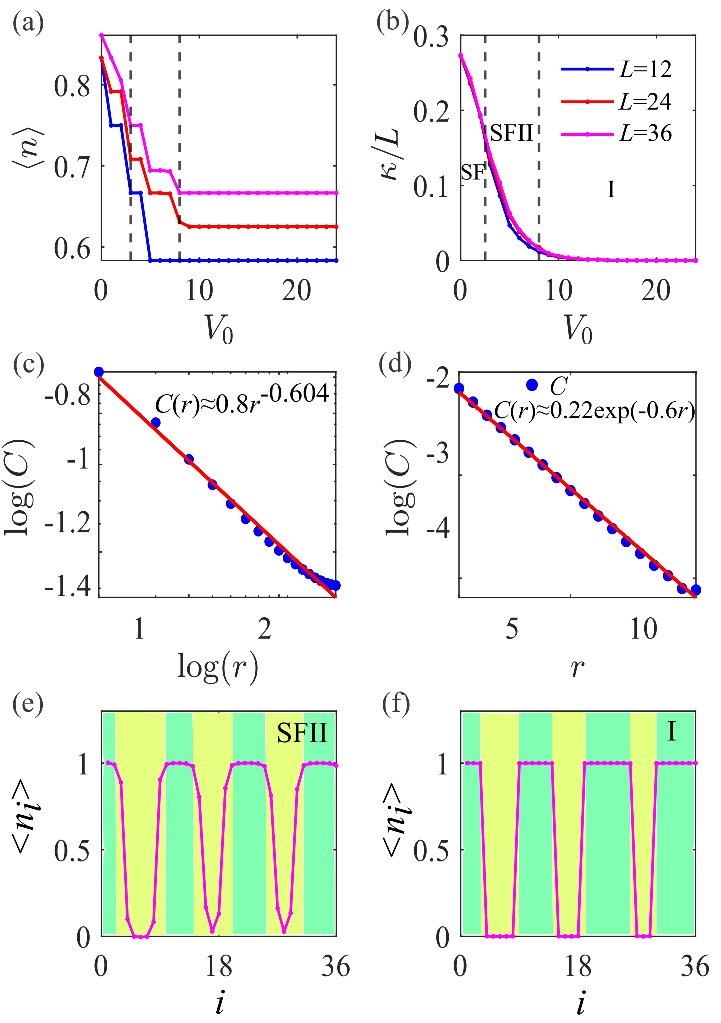}
\caption{
Phase characteristics of 1D lattices with $U=1$, $\mu=0.6$, and $V=0$.
(a) $\langle n \rangle$ vs $V_0$ (fixed $M_R=0.7$); (b) $\kappa/L$ vs $V_0$ for $L=12/24/36$.
(c) Algebraic decay of SF-phase correlations ($C(r)\approx0.8r^{-0.604}$); (d) exponential decay of SFII-phase correlations ($C(r)\approx0.22\exp(-0.6r)$).
(e,f) Real-space average density $\langle n_i \rangle$ for SFII and I phases (arrows mark parameters).
Physically, $V_0$ tunes phase transitions: increasing $V_0$ drives SF→SFII→I phase, with correlation decay switching from algebraic (SF)  to exponential (SFII and I).}
\label{fig:c}
\end{figure}

The method for distinguishing the SFII and I phases is defined as follows. An indicator is the particle number density $\langle n \rangle$, as shown in Fig.~\ref{fig:c} (b). When the depth of the well increases, if the potential well $V_0$ is deep enough, the density forms a plateau. At this point, the tunneling term in the system can be neglected, and the system is in the classical limit, where quantum fluctuations are completely absent. Furthermore, the expectation value $\langle n_i \rangle$ can be calculated exactly using the onsite minimum energy criterion.
 Meanwhile, $\langle n_i \rangle$  for each site is not the same, that is, the system enters phase I.

To further distinguish and identify the SFII phase from the I phase,  compressibility $\kappa$ serves as another critical criterion.
As illustrated in Fig.~\ref{fig:c}(c), the compressibility $\kappa/L$ satisfies $\kappa/L > 0$ in the SFII phase, while $\kappa = 0$ in the I phase.
A minor inconvenience is that the compressibility does not abruptly switch from nonzero to zero, but gradually transitions, making it difficult to determine the phase transition point. Therefore, we also use a threshold here to distinguish between zero- and nonzero values.
The I phase is identified when two conditions are satisfied: compressibility falls below 5\% of its maximum at \( V_0 = 0 \), and the average particle number density $\langle n \rangle$ forms on a non-integer plateau.

The transition between the SF and SFII phases is uniquely characterized by the
off-diagonal correlation function $C(r) = \langle a^\dagger_i a_{i+r} \rangle$,
which serves as the key order parameter for this phase transition.
The detailed correlation characteristics of distinct phases at $M_R=0.7$ can be illustrated via $C(r)$.
In the global SF phase, the off-diagonal correlation exhibits algebraic decay,
a well-established signature of long-range superfluid coherence \cite{hi1,hi2,Zhang2017}:
\begin{equation}
C(r) = \langle a_i^{\dagger} a_{i+r} \rangle.
\end{equation}
In contrast, the local SFII phase is distinguished by exponential decay of $C(r)$,
indicating suppressed long-range coherence.
As shown in Fig.~\ref{fig:c}, panel (d) presents $\log(C)$ versus $\log(r)$,
which displays nearly linear behavior for the SF phase, confirming a power-law decay $C \approx C_0 r^{-\eta}$,
with $\eta$ denoting the decay exponent.
Panel (e) plots $\log(C)$ against $r$, showing approximately linear trends for the SFII phase,
which demonstrates exponential decay $C \approx C_0 e^{-ar}$.
The corresponding parameter positions of these two regimes are marked by red and blue arrows in Fig.~\ref{fig:c} (a).
In many cases, visual inspection alone is insufficient to unambiguously distinguish
algebraic and exponential decay. To eliminate subjective judgment,
we further adopt the goodness of fit as a quantitative criterion.
The fitting quality is evaluated based on regression residuals,
i.e., the deviation between numerical data and theoretical decay functions.
A smaller residual value indicates a better fitting result,
which provides an objective and reliable criterion to differentiate the SF and SFII phases.

In terms of the property of the density distribution $\langle n_i \rangle$,  due to the influence of the moiré potential, $\langle n_i \rangle$ is nonuniform in the SF, SFII and I phases. For example, in Fig.~\ref{fig:c} (f), in the SFII phase for the 1D system, the non-uniform density distribution is shown. In the regions marked in yellow, the density exhibits non-integer values, whereas in the green-marked regions, the density shows integer values.
In Fig.~\ref{fig:c} (g), the density of the number of particles $\langle n_i \rangle$ for phase I exhibits a pattern of (000~1111~0000~11111$\cdots$), forming an integer string composed of alternating 0s and 1s. Due to the inhomogeneous moiré potential, the length of each string of 0s or 1s is not necessarily the same. This configuration mirrors the two-dimensional case, where disk-shaped domains with a density of 2 emerge against a background density of 1~\cite{Meng_2023}.

As shown in Fig.~\ref{fig:c}  (e)  and Fig.~\ref{fig:c} (f) , the number of lattice sites occupied by local domains varies across different spatial positions.
In the SFII phase, integer-density and non-integer-density regions are highlighted in yellow and green, respectively. The widths of the three yellow localized domains are not identical, containing 7, 5, and 5 lattice sites respectively.
In the I phase, by contrast, sharp step-like boundaries emerge between occupied and empty regions. Meanwhile, the width of each localized domain is also inhomogeneous, with sizes corresponding to 5, 4, and 3 lattice sites in sequence.

In conclusion, the emergence of the SFII phase and the non-uniform insulator I phase is attributed to the relatively large height difference ($M_R$) between the potential barriers and potential wells. There are no particles at the potential barriers, while particles exist at the potential wells and exhibit a local superfluid or insulating phase. The potential barriers prevent these local phases from connecting to form a global phase.

 \begin{figure}[b]
\includegraphics[width=0.48\textwidth]{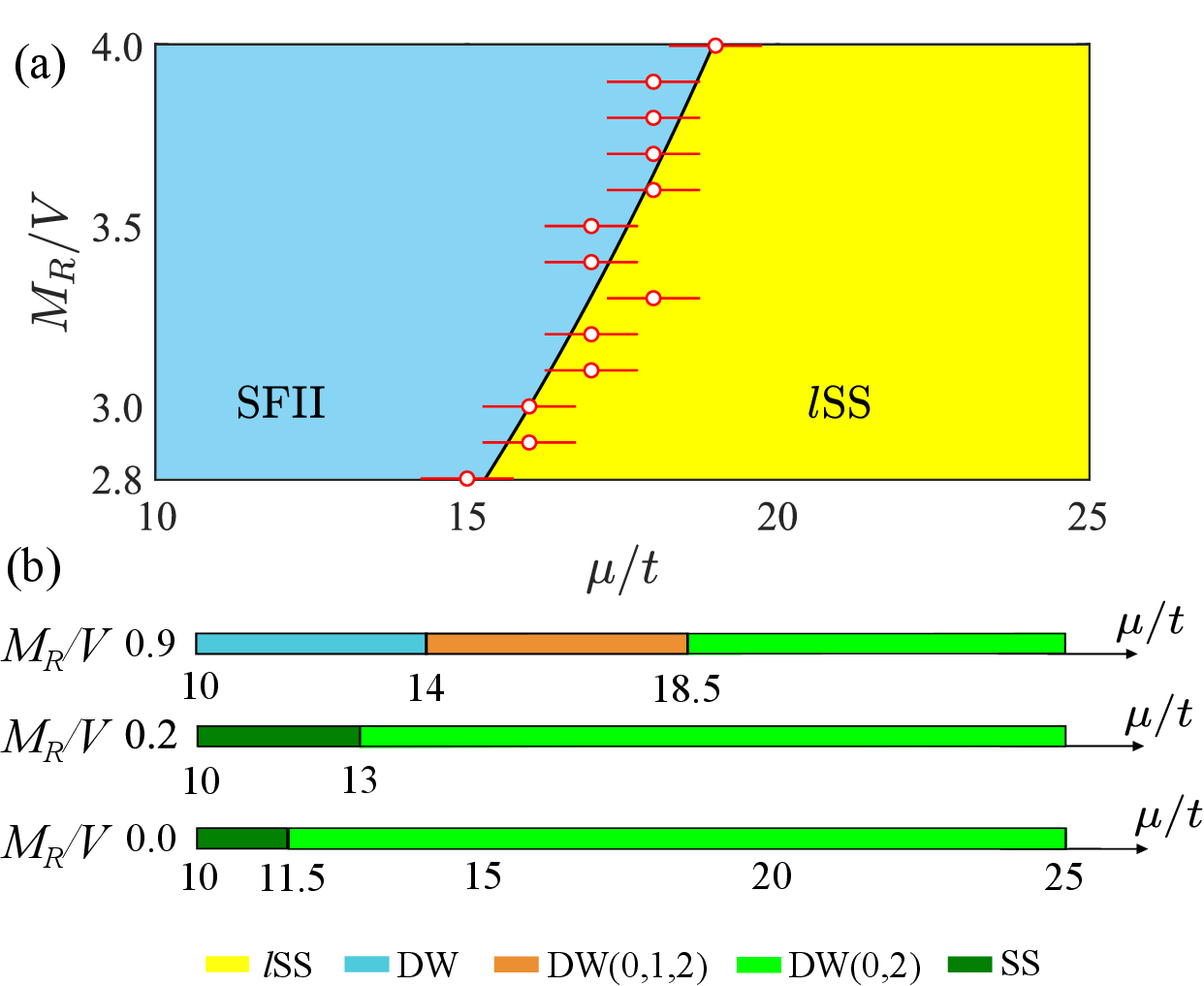}
\caption{Phase diagram of the one-dimensional system; the parameters used are $t/V=0.125$, $U/V=1.25$, and $V=1$.  (a) Under strong moiré potential, the two-dimensional phase diagram  shows the $l$SS phase (yellow region). As $\mu/t$ exceeds the approximate threshold of 15, enhanced repulsive interactions induce staggered local density order, driving a phase transition from SFII to $l$SS. The SFII-$l$SS boundary (red circles) rises with increasing $M_R/V$, demonstrating that larger modulation requires higher chemical potential to stabilize the $l$SS phase.
(b) In the weak moiré potential $0\le M_R/V\le1$, one-dimensional cuts at $M_R/V=0,0.2,0.9$  reveal SS, DW, DW(0,1,2), and DW(0,2) phases.
}
\label{fig:dwlss}
\end{figure}
\subsection{Supersolid and local supersolid in an interacting system ($V\ne 0$)}
Figure \ref{fig:dwlss} (a) presents the phase diagram of the one-dimensional system at large $M_R/V$ values, where the $l$SS phase emerges as the dominant phase in the yellow region. The phase boundaries are denoted by red circles with error bars and and the solid line is a guide to the eye.
For $M_R/V > 2.8$, the system is characterized by high potential barriers and numerous localized regions, giving rise to rich local quantum phases.
At small chemical potential $\mu/t$ (e.g., $\mu/t < 15$), the particle density in local regions is low, and the repulsive interaction $V$ (fixed at $V=1$ in this phase diagram) plays a weak role; accordingly, the system remains in the SFII phase (blue region).
As $\mu/t$ exceeds the approximate threshold of 15, more particles fill the local domains, enhancing the repulsive interaction $V$ (with $V=1$) to become the dominant interaction.
This strong repulsion induces a staggered particle density arrangement in the local regions, driving the system to undergo a phase transition from the SFII phase to the $l$SS phase.
Notably, the phase boundary between SFII and $l$SS (red circle markers) shows a clear upward trend with increasing $M_R/V$, indicating that higher potential modulation $M_R/V$ requires a larger chemical potential to stabilize the $l$SS phase.

For comparison, the lower panel, i.e., Fig.\ref{fig:dwlss}(b), illustrates the phase evolution at small $M_R/V$ ($0 \leq M_R/V \leq 1$), where the system exhibits conventional phases including SS, DW, DW(0,1,2), and DW(0,2) instead of local phases—further highlighting that the $l$SS phase is a unique feature of strong moiré potential modulation (large $M_R/V$).

\begin{figure}[htb]
  \centering
\includegraphics[width=\linewidth]{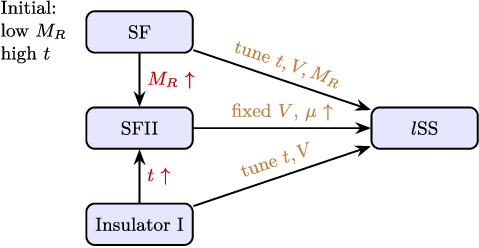}
  \caption{Schematic of the phase evolution pathways leading to the  $l$SS phase. The diagram shows that the $l$SS phase can be accessed from the SF phase, the SFII phase, and insulating I phases via different control parameters.
}
  \label{fig:m}
  \end{figure}

Even though the primary focus of this work is the $l$SS state induced by strong moiré potentials, the exploration of the SF, SFII, MI, and I phases is indispensable, since the $l$SS emerges from these phases and inherits key physical properties from them. As illustrated in the phase diagram of Fig.~\ref{fig:dwlss}, the $l$SS phase can be accessed directly from the SFII phase by increasing the particle number (chemical potential $\mu$) at fixed interaction strength $V$. Meanwhile, as shown in Fig.~\ref{fig:c}, the SFII phase itself emerges from two distinct routes: it can be reached from the SF phase by increasing the moiré potential strength $M_R$, or from the insulating I phase by increasing the hopping amplitude $t$ (decreasing $V_0$). These pathways as shown in Fig.~\ref{fig:m} clearly demonstrate that the $l$SS phase is deeply connected to its neighboring SF, SFII, and insulating phases, making a comprehensive understanding of these parent phases essential for characterizing the formation and stability of the $l$SS state. While these phases have been mentioned in previous experimental work~\cite{Meng_2023}, a systematic and high-precision numerical investigation using methods such as DMRG is still lacking. This motivates the present study to explore the full phase diagram, including the SF, SFII, and insulating phases, to establish a complete and consistent microscopic picture of the phase transitions leading to the $l$SS phase.

\subsubsection{The SS phase at small $M_R$}
In terms of parameter selection, to avoid blind parameter searching, we build on the extended soft-core BH model in Ref.~\cite{ss1d}.
The phase diagram was shown in plane ($\mu/t$, $t/V$) with the fixed parameters $U=10t$ and $V=1$.  In the range $10<\mu/t<12$, the stable SS phase was found to exist.  In this paper, we limit our discussion of the phase diagram to the region where $\mu/t > 10$.%

As shown in Fig~\ref{fig:dwlss}, when $M_R = 0$, along the $\mu/t$ axis ranging from 10 to 12, the phase diagram has an SS phase indicated by the dark green line segment, consistent with the result from Ref.~\cite{ss1d}.
Furthermore, when $\mu/t>12$, a phase DW emerges with a pattern of $020202\cdots$ labeled as DW(0,2).
The SS phase can be considered to be induced by hole excitations in the DW(0,2) phase.
When $M_R = 0.2$, the SS phase still exists; however, the particle number density distribution $\langle n_{i}\rangle$ already shows obvious modulation by the moiré potential.
However, from the perspective of either the particle number distribution or the tunneling correlation \(\langle a_i^{\dagger} a_{i+1}\rangle\), the entire lattice does not yet appear to be divided into distinct region. For brevity, we still label it as SS.

When $M_R = 0.9$, no SS phase is observed. This is because a large $M_R$ destroys the off-diagonal long-range order, and the system instead exhibits various density-wave phases. In some cases, the particle number density is non-integer—arising from the varying values of the local potential $-(\mu - M_i)n_i$—yet the expectation value of the particle number $\langle n_i \rangle$ still follows an alternating arrangement; we refer to this state as a DW phase. If the effect of  \(M_R\) is negligible, the system can exhibit integer density filling, DW(0,1,2) and DW(0,2) indicate that the particle numbers are distinct integers or close to integers.
 Increasing $M_R$ to 2.8, $l$SS emerges. In the thermodynamic limit, the global structure factor equals zero, while the local structure factor is non-zero.

\begin{figure}[b]
\centering
\includegraphics[width=0.5\textwidth, height=0.65\textwidth]{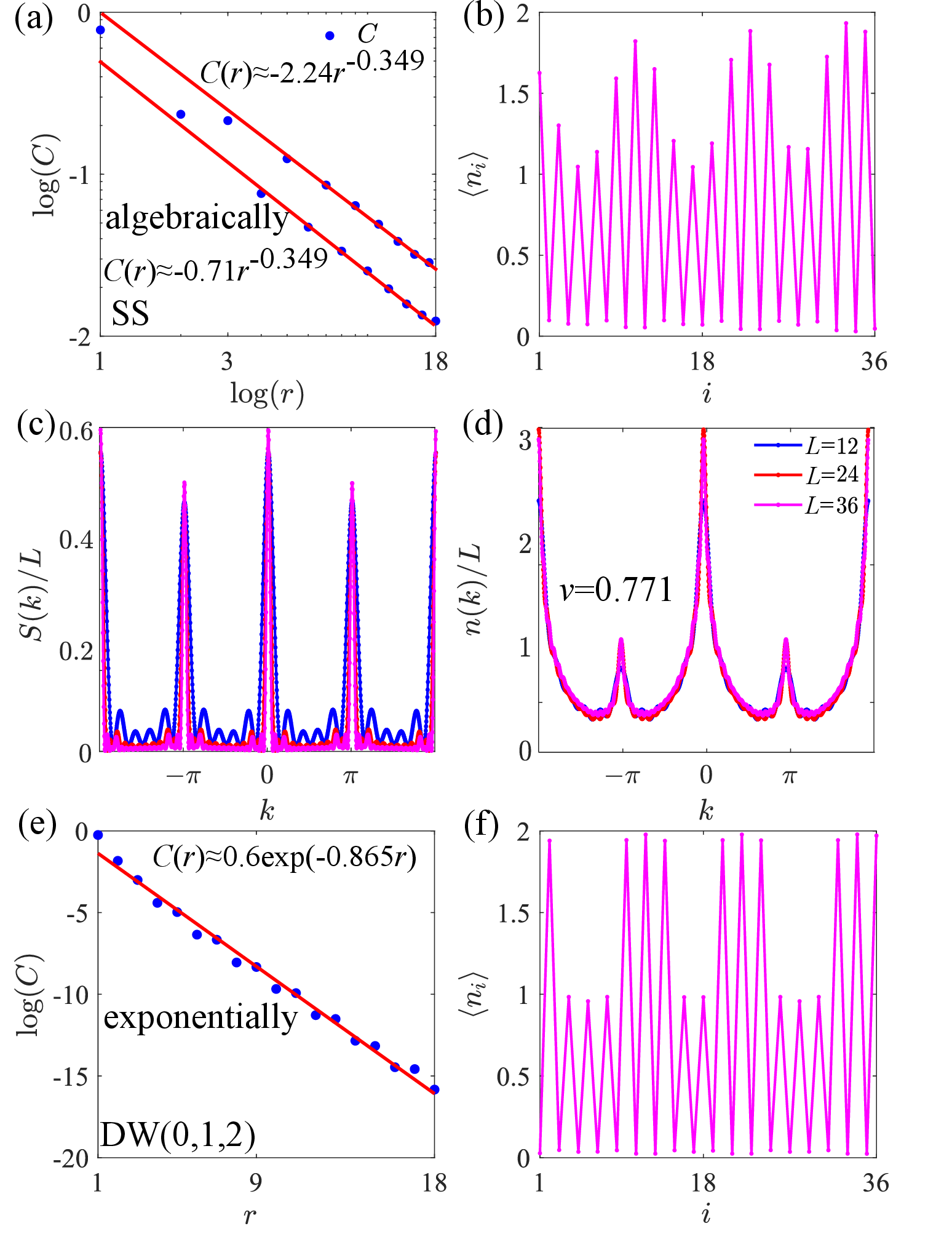}
\caption{Detailed characterization of the SS and DW$(0,1,2)$  phases in the one-dimensional lattice model.
For the SS phase:
(a) The correlation function $C(r)$ exhibits algebraic decay, $C(r)= r^{-\eta}$ with $\eta\approx0.349$, consistent with quasi-long-range order;
(b) Spatial distribution of the particle number density $\langle n_i\rangle$, showing a slow modulation envelope induced by the moiré potential, with a local alternating pattern of occupied/empty sites characteristic of the underlying DW order;
(c) Structure factor $S(k)/L$, with sharp peaks at wavevectors $k=\pm \pi$;
(d) Momentum distribution $n(k)/L$ for system sizes $L=12,24,36$, showing a peaks at $k=0$ and $\pm \pi$ with visialbility $\nu=0.771$.
For the DW$(0,1,2)$ phase:
(e) The correlation function $C(r)$ decays exponentially, $C(r)=\exp(-ar)$ with correlation length $a\approx 0.865$, indicating short-range order;
(f) Spatial distribution of the particle density $\langle n_i\rangle$, showing a periodic density-modulated pattern characteristic of the density wave phase. }
\label{fig:xijie30}
\end{figure}

The careful analysis of the SS  and DW(0,1,2) phases is shown in Fig.~\ref{fig:xijie30}. In Fig.~\ref{fig:xijie30}(a),  with parameters $M_R = 0.2$ and $\mu/t = 11$, for the SS phase, the correlation function $C(r)$ is plotted for the SF order in the SS phase.
Due to the oscillations in the correlation function, we fit the local maxima and minima of the oscillations separately, and find $C(r)$ obeys the way of algebraic decay, i.e., \begin{equation}
\begin{aligned}
    C(r) &= -0.71 \, r^{-0.349}, \\
    C(r) &= -2.24 \, r^{-0.349}
\end{aligned}
\end{equation}
which means that the system exhibits off-diagonal quasi-long-range order, an indicator of the SF order in the SS phase. The solid order of the SS phase can be observed from the staggered pattern in the distribution of the particle number expectation value \(\langle n_i \rangle\).
In Fig.~\ref{fig:xijie30}(b), the spatial density distribution for the SS phase is illustrated, with data shown for a system size of $L=36$. The particle number density displays non-uniform, non-integer, however spatially oscillatory behavior means that it has properties of solid.
The structural factor $S({k})/L$ in Fig.~\ref{fig:xijie30}(c), exhibit sharp peaks near \( k = \pm \pi \). As the system size increases, the values of these peaks remain present and stable, indicating that this solid order exists stably in the thermodynamic limit.

 In
Fig.~\ref{fig:xijie30}(d), for the SS phase, momentum distribution $n(k)$ shows a sharp peak in at $k=\pi$ and $k=0$, with the corresponding visibility  $v$ labeled as 0.771.
However, for the $l$SS phase, it is difficult to distinguish its key features using the global momentum distribution $n(k)$. Since the $l$SS phase lacks global phase coherence (evidenced by the exponential decay of correlations, see Fig.~\ref{fig:lssxin}(a)), the interference peaks in $n(k)$ are significantly broadened or fully suppressed, in  contrast to the sharp peaks observed in the global SS and global SF phases.
This broadening makes it difficult to unambiguously characterize the coherent nature of the $l$SS phase using $n(k)$.
If we instead define a \textit{local} momentum distribution $n_c(k)$ (in analogy with the local structure factor), i.e., the Fourier transform of the in-island correlation function, it similarly shows no sharp peaks in momentum space.
This is because the global correlation function decays exponentially, which we have explicitly verified in our calculations.

Another interesting phase is the emerged DW(0,1,2) phase due to the non-uniform moiré potential.
With parameters $M_R=0.9$ and $\mu/t=15$, the  correlation function $C(r)$ exhibits exponential decay as shown in  Fig.~\ref{fig:xijie30}(e) for this DW phase.
 The density profile in Fig.~\ref{fig:xijie30}(f) represents the configuration. It is formed by inserting zeros into the density pattern of the non-uniform insulator reported in Ref.~\cite{Meng_2023}. Specifically, in the absence of the nearest neighborhood interaction $V$, the particle arrangement follows the sequence $222221111222\cdots$. When a finite $V$ is introduced into the Hamiltonian, the configuration evolves into
$2020201010\cdots$ to minimize the total energy.

\subsubsection{ The $l$SS phase at larger $M_R$}
\begin{figure}[t]
\includegraphics[width=0.45\textwidth,height=0.5\textwidth]{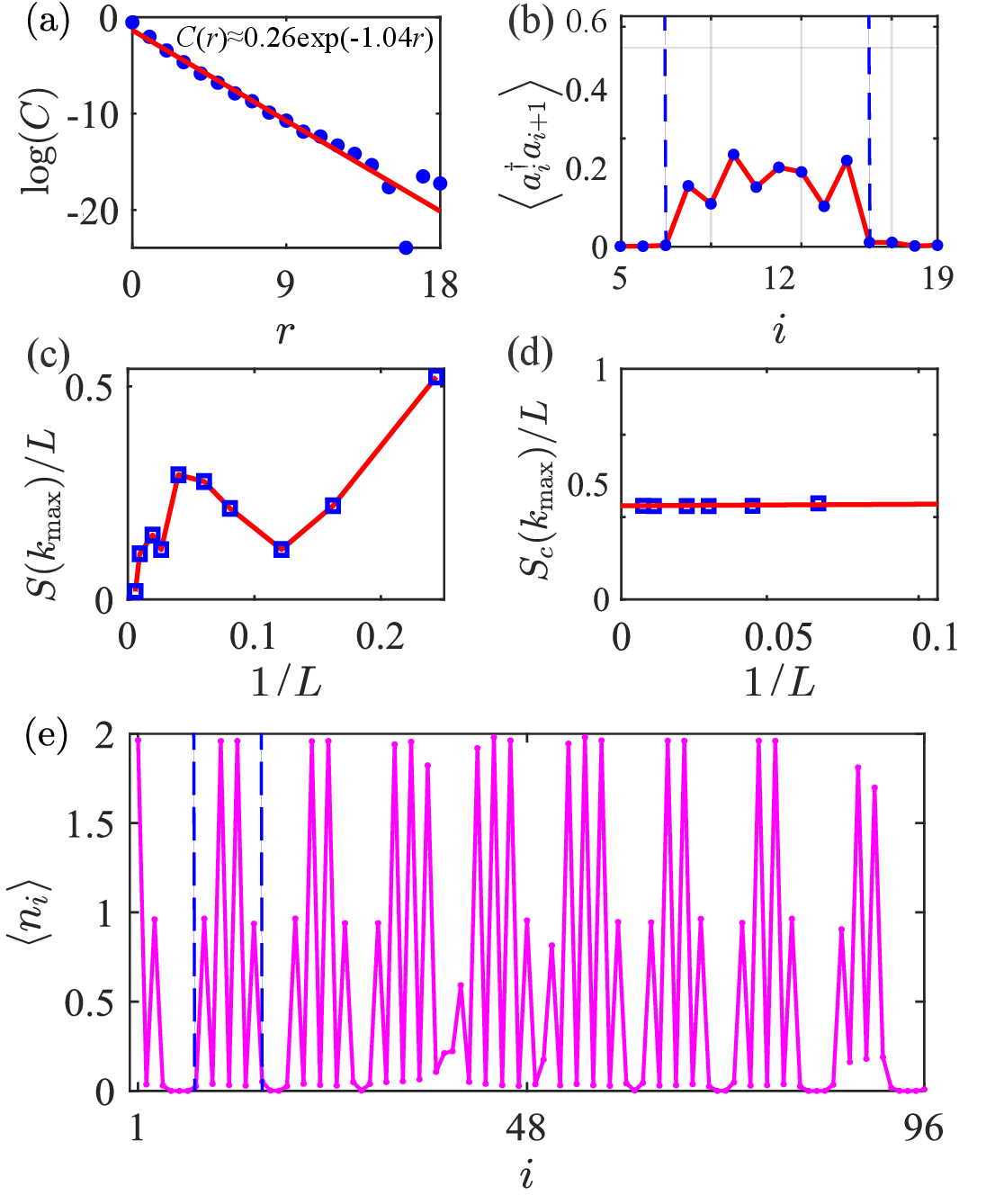}
\caption{Local SS phase details:
(a) Global off-diagonal correlation function $C(r)$, showing exponential decay $C(r)\approx 0.26\exp(-1.04r)$, indicating the absence of global phase coherence.
(b) Local correlation $\langle a_i^\dagger a_{i+1}\rangle$ within the region $8\le i\le 16$, showing non-zero values that demonstrate local superfluid coherence.
(c) Finite-size scaling of the global structure factor $S(k_{\text{max}})/L$, which tends to zero in the thermodynamic limit ($L\to\infty$), confirming the absence of long-range superfluid order. 
(d) Finite-size scaling of the local structure factor $S_c(k_{\text{max}})/L$, which remains non-zero in the thermodynamic limit, indicating persistent local superfluid correlations.
Here $L$ ranges from 4 to 144 in (c) and (d) are included in the scaling analysis.
(e) Spatial distribution of particle density $\langle n_i\rangle$, exhibiting a staggered local density pattern within the moiré unit cells (highlighted by dashed lines), consistent with the coexistence of local superfluidity and density wave order.
}
\label{fig:lssxin}
\end{figure}

To distinguish the $l$SS phase from the normal SS phase, two indicators can be used.  In the $l$SS phase, the long-range
off-diagonal correlation function decays exponentially [see Fig.~\ref{fig:lssxin} (a)], and the local expectation values $C(1)=\langle a_ia_{i+1}\rangle$ are nonzero [see Fig.~\ref{fig:lssxin} (b)] in the local regimes.
The lattice sites between the two blue dashed lines are chosen as the boundaries of different local regions.
In the local potential well region $8\le i \le15$, the off-diagonal correlation $\langle a_i^\dagger a_{i+1} \rangle$ is nonzero.
In the local potential barrier regions $5\le i \le7$ and $16\le i \le19$, the off-diagonal correlation $\langle a_i^\dagger a_{i+1} \rangle$ vanishes.

Furthermore, in the $l$SS phase, the global structure factor exhibits a peak near
$k=\pi$ that scales to zero in the thermodynamic limit [see Fig.~\ref{fig:lssxin} (c)].
Due to the particularity and inhomogeneity of the moiré potential, we observe that the structure factor does not decrease monotonically with the inverse of size $1/L$.
This is mainly due to the effect of the incommensurate moiré potential. In some potential wells, sublattice sites with even coordinates are occupied (with a particle number of 1), while those with odd coordinates are unoccupied (with a particle number of 0). In other potential wells of the lattice, however, the occupancy is swapped. These inconsistencies lead to the vanishing of the structure factor of the entire lattice in the thermodynamic limit.
This is quite different from a homogeneous system.

If we focus on the density or correlation within just one potential well (e.g., the region bounded by the two dashed lines in Fig.~\ref{fig:lssxin} (e)), the local structure factor remains non-zero [see Fig.~\ref{fig:lssxin} (d)].
$k_{\text{max}}$ in the label $S(k_{\text{max}})$ indicates that the positions of the peaks lie around $k=\pi$ rather than exactly at $k=\pi$.

\begin{figure}[b]
\includegraphics[width=0.5\textwidth]{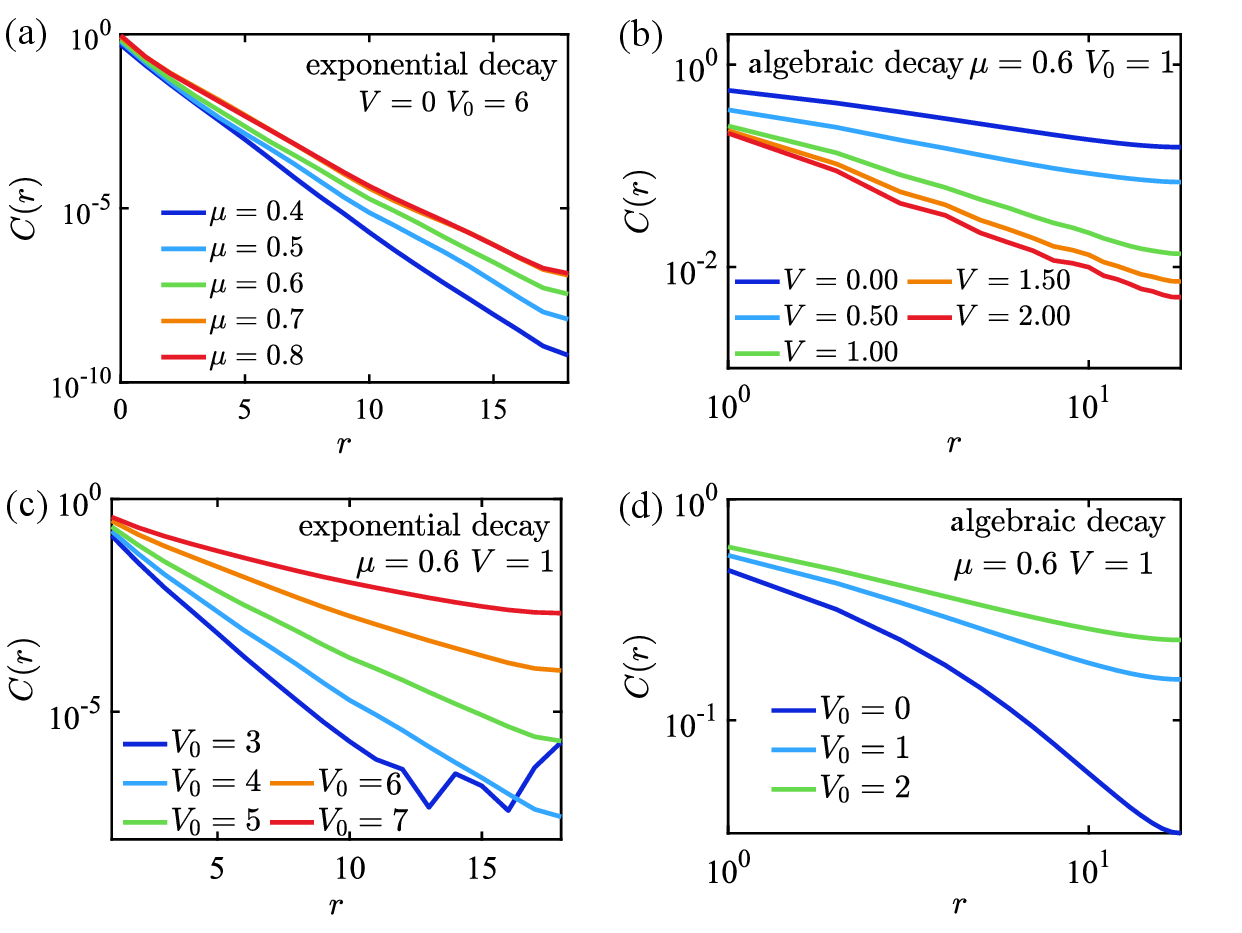}
\caption{Parameter dependence of off-diagonal correlation $C(r)$ decay (exponents in Table~\ref{tab:slope_fit}).
(a) Semi-log plot of $C(r)$ ($V=0$, $V_0=6$, $\mu=0.4$–$0.8$): exponential decay $C(r)\approx C_0 e^{-ar}$, $a$ from $-1.15(2)$ to $-0.87(3)$ (slower decay with larger $\mu$).
(b) Double-log plot of $C(r)$ ($\mu=0.6$, $V_0=1$, $V=0$–$2.0$): faster exponential decay with $a$ from $-0.47(2)$ to $-1.31(3)$ (SF→SS phase).
(c) Semi-log plot of $C(r)$ ($\mu=0.6$, $V=1$, $V_0=3.0$–$7.0$): faster exponential decay with larger $V_0$ (smaller $t$).
(d) Double-log plot of $C(r)$ ($\mu=0.6$, $V=1$, $V_0=0$–$2.0$): algebraic decay, accelerating with larger $V_0$.}
\label{fig:sigetu}
\end{figure}

\begin{table}[t]
  \centering
  \caption{Fitted slopes (scaling exponents) of correlation function decay under different parameters with fixed $M_R=0.7$}
  \label{tab:slope_fit}
  \begin{tabular}{cccc}
    \toprule
    Subplot & Fixed parameter  & Variable parameter & Fitted slope  \\
    (a) & $V=0,\ V_0=6$ & $\mu=0.4$ & $-1.15(2)$ \\
        &               & $\mu=0.5$ & $-1.01(3)$ \\
        &    exponential decay           & $\mu=0.6$ & $-0.92(3)$ \\
        &               & $\mu=0.7$ & $-0.87(3)$ \\
        &               & $\mu=0.8$ & $-0.87(3)$ \\
    (b) & $\mu=0.6,\ V_0=1$ & $V=0.00$ & $-0.47(2)$ \\
        &                  & $V=0.50$ & $-1.00(3)$ \\
        &       algebraic decay             & $V=1.00$ & $-1.08(3)$ \\
        &                   & $V=1.50$ & $-1.25(3)$ \\
        &                   & $V=2.00$ & $-1.31(3)$ \\
    (c) & $\mu=0.6,\ V=1$ & $V_0=3$ & $-1.23(4)$ \\
        &                 & $V_0=4$ & $-0.63(1)$ \\
        &     exponential decay            & $V_0=5$ & $-0.72(3)$ \\
        &                 & $V_0=6$ & $-0.54(2)$ \\
        &                 & $V_0=7$ & $-0.36(2)$ \\
    (d) & $\mu=0.6,\ V=1$ & $V_0=0$ & $-0.20(3)$ \\
        &                 & $V_0=1$ & $-0.09(3)$ \\
        &   algebraic decay              & $V_0=2$ & $-0.07(2)$ \\
        \hline
        \hline
  \end{tabular}
\end{table}

\subsection{The decay exponents as a function of model parameters such as $\mu$ or $V_0(t)$ or $V$ }
In Fig.~\ref{fig:sigetu}, we systematically analyze the decay behavior of the off-diagonal correlation function $C(r)$ under varying parameters. The corresponding decay exponents for all panels are summarized in Table~\ref{tab:slope_fit}.
Overall, the correlation decay slows down with increasing $\mu$, while the decay becomes faster with larger $V$ or larger $V_0$ (i.e., smaller tunneling $t$).

Figure ~\ref{fig:sigetu}(a) presents $C(r)$ on a semi-logarithmic scale for $V=0$, $V_0=6$, and $\mu=0.4$ to $0.8$.
The linear profiles confirm exponential decay $C \approx C_0 e^{-ar}$, with the decay exponent $a$ decreasing from $-1.15(2)$ to $-0.87(3)$ as $\mu$ increases.
As $\mu$ increases, the particle density rises accordingly, leading to a slower correlation decay, which is physically reasonable.
At the initial value of $\mu$, the local barrier region is nearly empty, similar to the particle configuration in Fig.~\ref{fig:c} (e).
The local barrier region is nearly empty with very low particle density.
As $\mu$ increases, the particle density in the barrier region gradually increases, and eventually evolves into a obvious local superfluid state.
Consequently, the off-diagonal correlation decays much more slowly.

Figure ~\ref{fig:sigetu}(b) shows $C(r)$ for fixed $\mu=0.6$, $V_0=1$, and $V=0$ to $2.0$ on a double-logarithmic scale.
The exponential decay accelerates with increasing $V$, with $a$ varying from $-0.47(2)$ to $-1.31(3)$, indicating weakened off-diagonal order under stronger modulation. At the same time, the system evolves from the SF phase into the SS phase.
Figure ~\ref{fig:sigetu}(c) displays $C(r)$ for fixed $\mu=0.6$, $V=1$, and $V_0=3.0$ to $7.0$.
The correlation exhibits exponential decay and becomes faster as $V_0$ (i.e., tunneling $t$) increases.
Figure ~\ref{fig:sigetu}(d) shows $C(r)$ for fixed $\mu=0.6$, $V=1$, and $V_0=0$ to $2.0$.
In contrast to other panels, $C(r)$ shows algebraic decay, and the decay accelerates with increasing $V_0$ (decreasing $t$).

\subsection{Local phases under the moiré potential at larger system sizes}
\label{sec:Region}

\begin{figure}[t]
\centering
\includegraphics[width=1\linewidth]{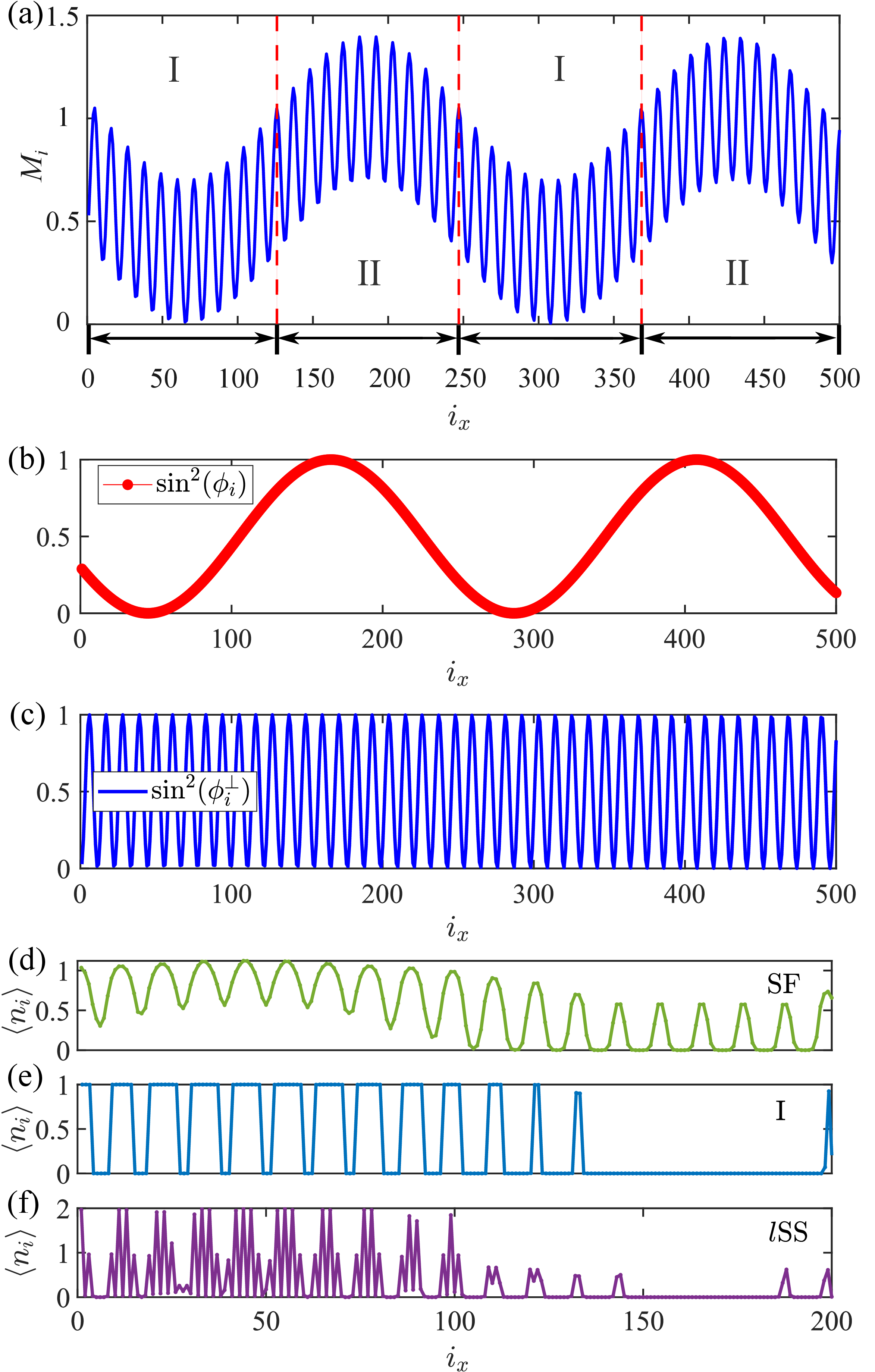}
\caption{Spatial structure of the moiré potential and particle density in large-scale systems.
(a) The moiré potential profile $M_i$ over a lattice of length $L=500$, showing periodic alternation between high-potential ``Region I" and low-potential ``Region II", with red dashed lines marking the boundaries of individual moiré supercells.
(b) The slow-modulation component $\sin^2(\phi_i)$ of the moiré potential, which drives the large-scale envelope oscillation.
(c) The fast-oscillating component $\sin^2(\phi_i^\perp)$ of the moiré potential, responsible for the short-wavelength fluctuations within each supercell.
(d)-(f) Particle number density distributions $\langle n_i\rangle$ for $L=200$ in three phases: (d) SF phase, (e) I phase, and (f) $l$SS phase. The SF and I phases show non-local and fully localized density patterns, respectively, while the $l$SS phase exhibits coexisting local density modulations and decaying correlations. All three phases remain stable in the region $i_x<100$, consistent with previous results at $L=96$.}
\label{fig:nx500muhsite}
\end{figure}

In previous results, for example Fig.~\ref{fig:lssxin} (e), we restrict the lattice size to $L=96$, and it is unclear how the states of our system would evolve for larger system sizes. Therefore, we plot the moir\'e potential profile for $L=500$ in Fig.~\ref{fig:nx500muhsite}(a) and observe an interesting phenomenon. Besides the short-period oscillations inside the moir\'e potential, a much longer-period oscillation is also present. This is related to the expression of the moir\'e potential given in Sec.~\ref{sec:model}:
\begin{align}
    M_i &= M_R \left[ \sin^2(\phi_i) + \sin^2(\phi_i^\perp) \right], \\
    \phi_i &= i_x \pi \cos\theta + i_y \pi \sin\theta, \\
    \phi_i^\perp &= i_y \pi \cos\theta - i_x \pi \sin\theta,
\end{align}
where the first term exhibits slow oscillations of $\sin^2(\phi_i)$, and the second term exhibits fast oscillations of $\sin^2(\phi_i^\perp)$, as shown in Figs.~\ref{fig:nx500muhsite}(b) and (c). This arises because at $\theta=5.21^{\circ}$, the coefficient of $i_x$ (i.e., $\cos\theta$) in the first term is much larger than that in the second term (i.e., $\sin\theta$). The first term corresponds to long-wavelength oscillations, while the second term corresponds to short-wavelength oscillations.

To verify whether the I, SF, and $l$SS phases identified in our previous  sections are stable at larger system sizes, we also calculate the particle number density distributions of the three phases (SF, I, and $l$SS) for $L=200$. For reproducibility, the corresponding parameters are provided: the SF phase is obtained with $V_0=1$, $\mu=0.6$, $U=1$, and $M_R=0.7$; the I phase is obtained with $V_0=15$, $\mu=0.6$, $U=1$, and $M_R=0.9$; and the $l$SS phase is obtained with $t=0.125$, $\mu=2.625$, $U=1.25$, and $M_R=2.8$.

In Figs.~\ref{fig:nx500muhsite}(d), (e) and (f), it can be seen from the results that the particle number at each lattice site decreases significantly for $i_x>100$, which is caused by the higher moir\'e potential in region II. For $i_x<100$, the particle number at each position of the $L=200$ system remains consistent with that of the $L=96$ system, which does not affect our previous phase diagram and conclusions.
\subsection{The role of angle $\theta$}
In this section, we study the role of the angle. First, we present the angular derivation formula to demonstrate the theoretical origin of $\theta = 5.21^{\circ}$. Second, we plot the corresponding  moir\'e potential for smaller and larger commensurate angles $\theta$.
In brief, the fixed angle $\theta=5.21^\circ$ is a typical commensurate twist angle derived from Pythagorean triples, which guarantees a well-defined periodic moiré superlattice.
Changing the angle significantly modifies the moiré supercell size:
too small angles lead to extremely large supercells (numerically intractable), while too large angles yield overly small supercells (unable to host localized phases).
For incommensurate angles, the periodic moiré pattern is destroyed, resulting in a disordered potential.
All details, derivations, and demonstrations can be found  as follows.

\begin{figure}[htbp]
  \centering
  \includegraphics[width=0.9\linewidth]{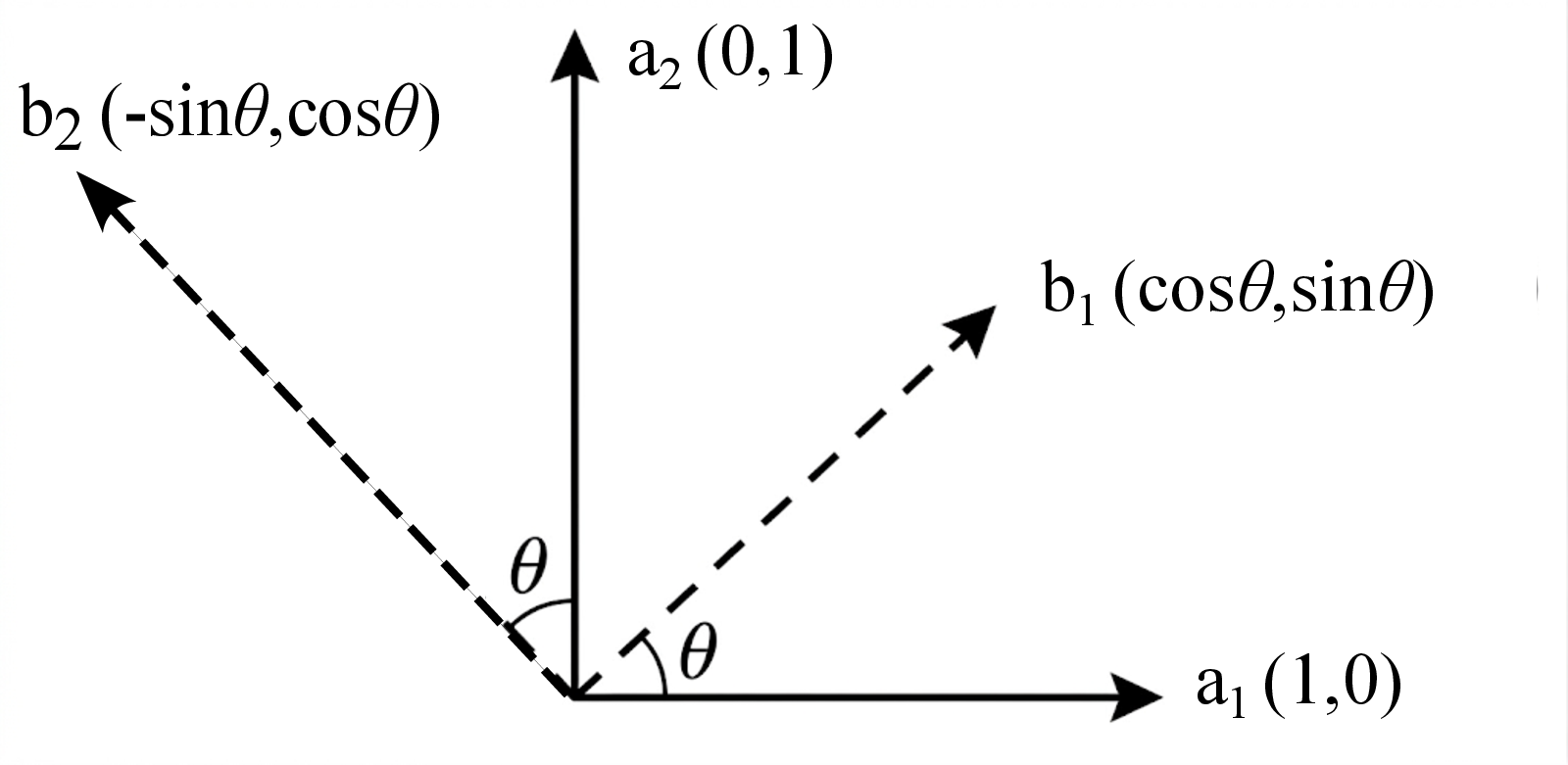}
 \caption{Definition of lattice basis vectors for the bottom (unrotated) and rotated top layers, showing the basis vectors $\mathbf{a}_1, \mathbf{a}_2$ of the original square lattice and $\mathbf{b}_1, \mathbf{b}_2$ of the layer rotated counterclockwise by an angle $\theta$. This geometric construction underpins the derivation of the moiré superlattice commensurability condition.}
  \label{fig:b1}
\end{figure}

Consider an unrotated, perfect square lattice as the bottom layer with basis vectors $\mathbf{a}_1 = (1, 0)$ and $\mathbf{a}_2 = (0, 1)$ as shown in Fig.~\ref{fig:b1}. Any lattice point in this layer can be represented by integer coordinates:
\begin{equation}
    \mathbf{R}_1 = x_1 \mathbf{a}_1 + y_1 \mathbf{a}_2 = (x_1, y_1),
\end{equation}
where $x_1, y_1 \in \mathbb{Z}$. When the top layer is rotated counter-clockwise by an angle $\theta$, its new basis vectors become $\mathbf{b}_1 = (\cos\theta, \sin\theta)$ and $\mathbf{b}_2 = (-\sin\theta, \cos\theta)$. Any lattice point in the rotated top layer, expressed in the original coordinate system, is given by:
\begin{equation}
    \mathbf{R}_2 = x_2 \mathbf{b}_1 + y_2 \mathbf{b}_2 = (x_2\cos\theta - y_2\sin\theta, \; x_2\sin\theta + y_2\cos\theta),
\end{equation}
where $x_2, y_2 \in \mathbb{Z}$.

The formation of a periodic moir\'e superlattice requires the existence of coincident points in space, such that $\mathbf{R}_1 = \mathbf{R}_2$. Expanding this into components yields:
\begin{align}
    x_1 &= x_2\cos\theta - y_2\sin\theta, \label{eq:x} \\
    y_1 &= x_2\sin\theta + y_2\cos\theta, \label{eq:y}
\end{align}
Since the coefficients $x_1, y_1, x_2, y_2$ must all be integers, it follows that the matrix elements of the rotation matrix, $\cos\theta$ and $\sin\theta$, must be rational numbers. Given that $\cos\theta$ and $\sin\theta$ are rational, they can be expressed as irreducible fractions:
\begin{equation}
    \cos\theta = \frac{A}{C}, \quad \sin\theta = \frac{B}{C},
\end{equation}
where $A, B, C$ are positive integers. Substituting these into the fundamental trigonometric identity $\cos^2\theta + \sin^2\theta = 1$ leads to:
\begin{equation}
    A^2 + B^2 = C^2.
\end{equation}
This indicates that the commensurability condition for a square lattice is mathematically equivalent to finding a set of Pythagorean triples—three positive integers that satisfy the sides of a right triangle.

All primitive Pythagorean triples $(A, B, C)$ can be generated by two coprime integers $m$ and $n$ (where $m > n > 0$) using the following general solution:
\begin{align}
    A &= m^2 - n^2 , \label{eq:A} \\
    B &= 2mn , \label{eq:B} \\
    C &= m^2 + n^2.\label{eq:C}
\end{align}
By substituting Eq. \eqref{eq:A} and \eqref{eq:C} back into the definition of $\cos\theta$, we rigorously derive the general formula for the commensurate angles of a square lattice:
\begin{equation}
    \cos\theta = \frac{m^2 - n^2}{m^2 + n^2}.
\end{equation}
It is worth noting that substituting Eq. \eqref{eq:B} into the definition of $\sin\theta$ yields
\begin{equation}
    \sin\theta = \frac{2mn}{m^2 + n^2}.
\end{equation}
Within the framework of direct derivation from the coordinate transformation matrix, the choice of $\frac{m^2 - n^2}{m^2 + n^2}$ as the analytical solution for the diagonal element $\cos\theta$ is more direct and rigorous. Furthermore, due to the high $C_4$ rotational symmetry of the square lattice, a rotation by a complementary angle $90^\circ - \theta$ yields a physically equivalent moir\'e pattern, where
\begin{equation}
   \cos(90^\circ - \theta) = \sin\theta = \frac{2mn}{m^2 + n^2}.
\end{equation}
This geometric equivalence perfectly explains why the latter formula often appears in the literature.

\begin{figure}[t]
  \centering
  \includegraphics[width=0.8\linewidth]{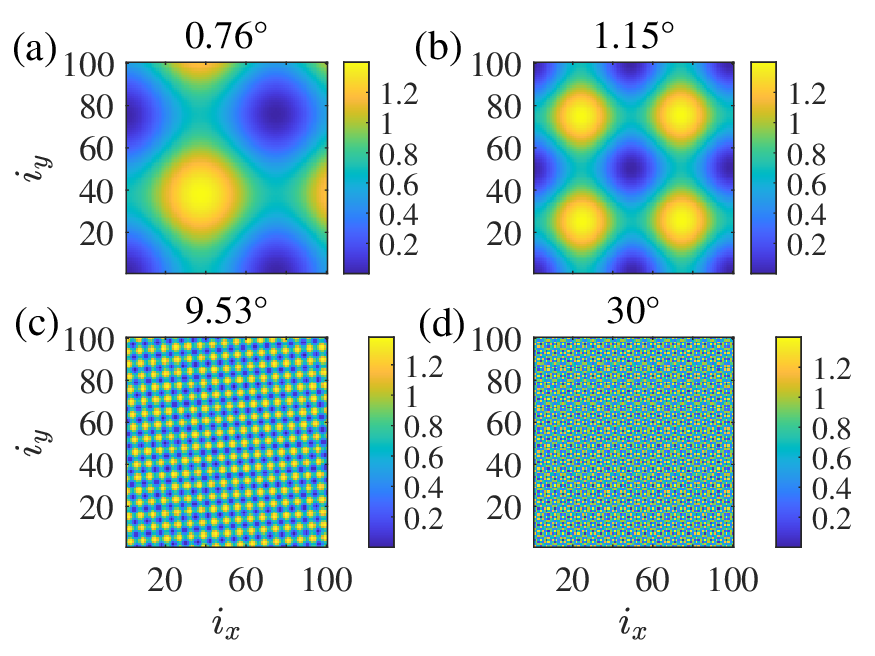}
  \caption{Spatial distributions of square moir\'e superlattices. (a,b) Extremely small commensurate twist angles of $\theta \approx 0.76^\circ$ ($m=150, n=1$) and $\theta \approx 1.15^\circ$ ($m=100, n=1$) yield exceptionally large supercell periods. (c) A larger commensurate angle of $\theta \approx 9.53^\circ$ ($m=12, n=1$) results in much smaller and denser supercells. The patterns in (a)-(c) explicitly verify the strict geometric relation $\cos\theta = (m^2-n^2)/(m^2+n^2)$, demonstrating that the twist angle inversely dictates the moir\'e supercell size. (d) At an incommensurate angle of $\theta = 30^\circ$ (where $\cos\theta = \sqrt{3}/2$ is irrational), the commensurate condition has no integer solutions. Consequently, the spatial periodicity is completely destroyed, reducing the effective potential to a quasi-random disordered landscape.}
  \label{fig:moire_angles}
\end{figure}

To intuitively demonstrate the impact of angle variation, we also present the moir\'e superlattice potentials for two alternative twist angles: two extremely small angles $\theta \approx 0.76^\circ$ ($m=150, n=1$), $\theta \approx 1.15^\circ$ corresponding to $m=100, n=1$, and a larger angle $\theta \approx 9.53^\circ$ corresponding to $m=12, n=1$. The lattice diagrams, both with dimensions of $100 \times 100$, are shown in Fig.~\ref{fig:moire_angles}. As illustrated, when the twist angle is too small ($\theta \approx 1.15^\circ$), the supercell spacing becomes excessively large (approaching 50 lattice sites), making it computationally prohibitive to simulate and observe the behavior across multiple supercells. Conversely, when the twist angle is too large ($\theta \approx 9.53^\circ$), the supercells become too small, hindering the observation of internal particle distribution differences required for complex local phases. Whether novel local quantum phases emerge at these extreme angles is left for future exploration.

Additionally, for incommensurate angles, the spatial periodicity is completely lost, and the effective potential reduces to a purely disordered landscape, as shown in Fig.~\ref{fig:moire_angles}(c).

\section{Discussion and Conclusions}
\label{sec:con}

Twisted bilayer systems have become a frontier for exploring emergent quantum phases, with theoretical work revealing that the twist angle can stabilize unique correlated states absent in single-layer or untwisted lattices~\cite{Zhang2025PRB,Ding2025PRA,Zeng2025} . Recent experiments by Ketterle’s group have realized a highly controllable bilayer system of ultracold atoms. While this realization employs untwisted  layers, the experimental architecture is fully compatible with introducing an additional inhomogeneous potential to emulate moiré superlattice physics~\cite{science.adh3023}. In this work, we do not simulate a bilayer system directly, but study a single-layer atomic system under an effective moiré potential, whose physics can be observed in each layer of realistic bilayer cold-atom setups~\cite{science.adh3023}.

In total, we numerically investigate the one-dimensional extended BH model with a moiré potential using the DMRG method, focusing on the microscopic characterization and identification criteria of the $l$SS phase. We show that the $l$SS phase emerges under strong moiré modulation and intermediate nearest-neighbor repulsion, and can be accessed from the SF, SFII, and I phases via tuning system parameters. The neighboring phases are essential to defining the formation pathways and stability of the $l$SS state.

A key result is the establishment of clear and experimentally observable signatures for the $l$SS phase. In contrast to a global supersolid, the $l$SS phase exhibits exponentially decaying global off-diagonal correlations and a vanishing global structure factor in the thermodynamic limit, while maintaining local superfluid coherence and local staggered density order within moiré supercells. These features unambiguously distinguish the $l$SS from the SFII phase and other conventional phases.

Although motivated by two-dimensional moiré experiments, our one-dimensional model enables high-precision numerical verification that is currently challenging in higher dimensions~\cite{Meng_2023,11}. In addition, all phase diagrams reported in this work are obtained at zero temperature, and the corresponding finite-temperature phase diagram remains to be explored. Our results provide a consistent microscopic picture of local quantum phases in moiré lattices and offer direct observables for future experimental detection of the $l$SS phase.

In conclusion, this work identifies the $l$SS as a well-defined quantum phase unique to strong moiré potentials, clarifies its characteristic fingerprints, and maps its evolution from neighboring phases. These findings deepen the understanding of correlated states in low-dimensional moiré systems and provide theoretical guidance for ultracold-atom experiments~\cite{Meng_2023,science.adh3023}.

{\it Acknowledgements:}  We would like to express our sincere gratitude to Professor Ming Gong and Professor Wei Han  for many useful discussions. We also thank the two anonymous referees for many helpful suggestions. W. Z. was supported by the Hefei National Research Center for Physical Sciences at the Microscale (Grant No. KF2021002) and the Fundamental Research Program of Shanxi Province (Grant Nos. 202303021221029, 202203021211337).
 Q. X. is supported from the Fundamental Research Program of Shanxi Province (Grant No. 20210302123152). Q.-Q. Shi is supported in part by the National Natural Science Foundation of China (Grant No. 12505003).

\appendix 

\begin{figure}[H]
  \centering
\includegraphics[width=0.48\textwidth]{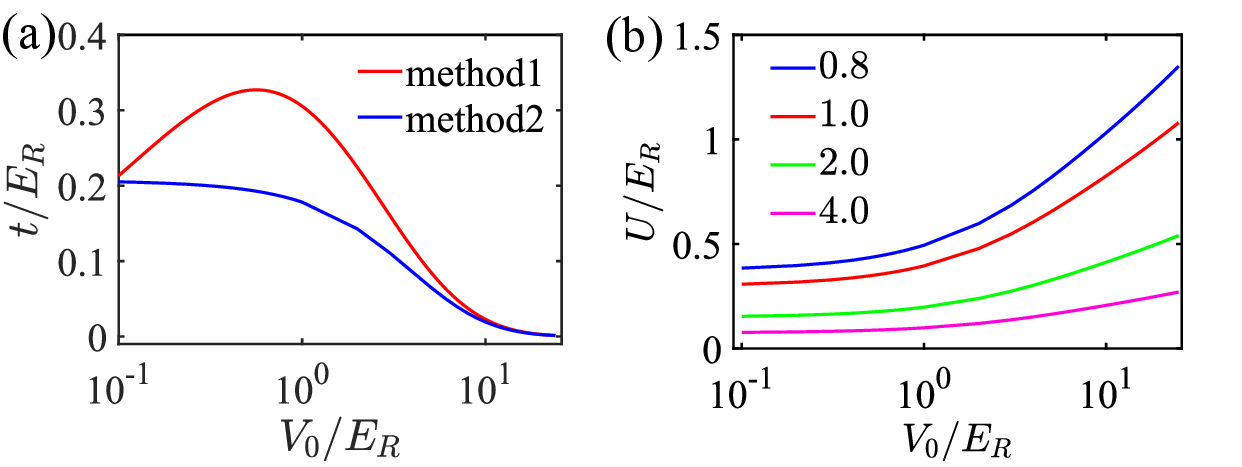} 
    \caption{(a) The hopping amplitude $t/E_R$ versus lattice depth $V_0/E_R$ calculated by two methods: 
method 1 (red) is the approximate analytical expression~\ref{eq:hopping}, 
and method 2 (blue) is the exact numerical result from Bloch wave function calculation. 
(b) On-site interaction strength $U/E_R$ versus $V_0/E_R$ for different interaction parameters $a$.}
    \label{fig:tU}
  \end{figure}
\section{Relationship between $t$, $U$ and $V_0$ from Bloch wave function calculation} 
\label{sec:app}

Since the phase diagram in Ref.~\cite{Meng_2023} uses $V_0$ as the horizontal axis, while our calculations adopt the tunneling amplitude $t$ as a parameter, a conversion between $V_0$ and $t$ is required. The approximate dependence of $t$ on $V_0$, given by Eq.~\eqref{eq:hopping} (first introduced in Section~\ref{sec:model}), is obtained by replacing the Wannier function with a Gaussian approximation~\cite{19}.
As shown in Fig.~\ref{fig:tU}, labeled as ``method 1", this approximate relation does not decrease monotonically, which is inconsistent with the physical expectation that deeper lattice potentials suppress tunneling. 

Therefore, we use an alternative method~\cite{github1} by Agdelma, as outlined below.
For a particle confined in an optical lattice, where the Hamiltonian is given by
\begin{equation}
H=\frac{p^{2}}{2 m}+V_{0} \sin ^{2} kx,
\end{equation}
where  $k=\pi / a_{0}$  with  $a_{0}$  the lattice spacing. 
According to Bloch's theorem, 
the wave function should be  \begin{equation}
\Phi_{q}^{(0)}(x)=\mathrm{e}^{i q x} u^{(0)}_q(x)=\mathrm{e}^{i q x} \sum_{\ell} C_{\ell}^{(0, q)} \mathrm{e}^{i 2 k \ell x},
\end{equation}
where $q$ is the vector on the reciprocal lattice, $C_{\ell}^{(0, q)}$ are the expansion coefficients for the lowest band.
The coefficients can be obtained by solving the eigenvalue equation as follows.
\begin{equation}
\sum_{\ell^{\prime}} H_{\ell, \ell^{\prime}} C_{\ell^{\prime}}^{(0, q)}=E_{q}^{(n)} C_{\ell}^{(0, q)},
\end{equation}
where $H$ is a tri-diagonal matrix and its elements should be 

\begin{equation}
H_{\ell, \ell^{\prime}}=\left\{\begin{array}{ll}
{\left[E_{R}\left(2 \ell+\frac{q}{k}\right)^{2}+\frac{V_{0}}{2}\right] \delta_{\ell, \ell^{\prime}}} & \text { if } \ell^{\prime}=\ell, \\
-\frac{V_{0}}{4}\left(\delta_{\ell, \ell^{\prime}+1}+\delta_{\ell, \ell^{\prime}-1}\right) & \text { if } \ell^{\prime} \neq \ell .
\end{array}\right.
\end{equation}
Here, the recoil energy $E_{R}$ is defined by
$E_{R}=\frac{\hbar^{2} k^{2}}{2 m}$. The Wannier function can be calculated directly by 
\begin{equation}
w\left(x-x_{j}\right)=\frac{1}{\sqrt{N a_{0}}} \sum_{q} \mathrm{e}^{i q x_{j}} \Phi_{q}^{(0)}(x).
\end{equation}
Finally, the tunneling $t$ is obtained by 
\begin{equation}
t=-\int_{-L / 2}^{L / 2} d x w^{*}(x)\left(-\frac{\hbar^{2}}{2 m} \frac{d^{2}}{d x^{2}}+V_{0} \sin ^{2} k x\right) w\left(x-a_{0}\right),
\end{equation} The code can be accessed at~\cite{github1} by Agdelma,
and the result is shown in Fig.~\ref{fig:tU} labeled as ``method 2".

In the contact interaction limit, the on-site interaction reduces to
\begin{equation}
U = g\int |w(x)|^4 dx,
\end{equation}
which corresponds to the zero-range $\delta$-function interaction between two atoms.
The coupling strength is given by $g=\frac{4}{a\pi^2}$, where $a$ is a tunable parameter with values $0.8$, $1.0$, $2.0$, and $4.0$, as shown in Fig.~\ref{fig:tU}(b).

\section{The effects of cut off $n_{max}$}
Here, we discuss the influence of the site occupation cutoff \(n_{\rm max}\) on the numerical results.
  For the $l$SS phases at with parameters $\mu = 2.375$, $M_R = 2.8$, $t = 0.125$, $U = 1.25$, increasing cutoff $n_{max}$  do not alter results.
  Similiarily, for DW phases, the  characteristics  remain  unchanged.
 This is because the on-site interaction is strong ($U/t = 10$), which suppresses multiple occupations.
However, it cannot be ensured that varying $n_{max}$ has no impact on the results within the SF phase. For this reason, we limit the scope of our work and omit further analysis of larger $n_{max}$ values. A comprehensive exploration of this phenomenon will be addressed in future research.

\bibliography{ref}
\end{document}